\documentclass[10pt,twocolumn]{article}

\usepackage{stmaryrd}
\usepackage{amsmath}
\usepackage{wrapfig}
\usepackage{amssymb}
\usepackage{amsthm}
\usepackage{enumerate}
\usepackage{multirow}
\usepackage{tikz}
\usetikzlibrary{arrows,decorations,decorations.shapes,decorations.markings,backgrounds,shapes,petri,topaths}
\usepackage{morehelp}
\usepackage{caption}
\usepackage[hypertexnames=false]{hyperref}
\usepackage{tkz-berge}
\usepackage[position=top]{subfig}

\newtheorem{theorem}{Theorem}[section]
\newtheorem{corollary}[theorem]{Corollary}
\newtheorem{lemma}[theorem]{Lemma}
\newtheorem{observation}[theorem]{Observation}

\newtheorem{proposition}[theorem]{Proposition}
\newtheorem{definition}{Definition}

\newtheorem{example}[theorem]{Example}

\DeclareMathOperator*{\argmin}{argmin}
\DeclareMathOperator*{\argmax}{argmax}

\begin{document}

\def\qedsymbol{$\square$}
\def\shortcite{\cite}
\newcommand{\ex}[2]{\mathbb{E}_{#1}{\left[#2\right]}}
\newcommand{\ind}[1]{\llbracket #1 \rrbracket}
\def\({\left(}
\def\){\right)}
\def\ol{\overline}
\def\H{\mathbf{H}}
\newcommand{\labeq}[2]{   
\begin{equation}
\label{eq:#1}
#2
\end{equation}}
\newcommand{\tup}[1]{\left\langle #1\right\rangle}
\renewcommand{\vec}[1]{\mathbf{#1}}
\def\mc{\mathcal}
\def\({\left(}
\def\){\right)}
\newcommand{\floor}[1]{\left\lfloor #1 \right\rfloor}
\newcommand{\ceil}[1]{\left\lceil #1 \right\rceil}
\def\eps{\epsilon}
\newcommand{\todo}[1]{\textbf{TODO:} #1 \textbf{::}}
\newenvironment{rtheorem}[1]{\indent\medskip\textsc{Theorem~\ref{#1}.}\begin{itshape}}{\end{itshape}}
\newenvironment{rprop}[1]{\indent\medskip\textsc{Proposition~\ref{#1}.}\begin{itshape}}{\end{itshape}}
\newenvironment{rlemma}[1]{\indent\medskip\textsc{Lemma~\ref{#1}.}\begin{itshape}}{\end{itshape}}
\newenvironment{aproof}[1]{\noindent\textit{#1.}}{\hfill\qed\vspace{1ex}}
\newenvironment{atheorem}[1]{\medskip\textsc{#1.}\begin{itshape}}{\end{itshape}\medskip}

\newcommand\subpar[1]{\medskip\vspace{-0mm}\noindent \textbf{#1.} }
\newcommand\newpar[1]{\vspace{-0mm}\paragraph{#1}}
\def\emptypar{\medskip\noindent}

\title{Equilibrium in Labor Markets  with Few Firms}
\author{
Reshef Meir\thanks{Microsoft Israel R\&D Center and Hebrew University of Jerusalem}
\and
Moshe Tennenholtz\thanks{Microsoft Israel R\&D Center and Technion-Israel Institute of Technology}\\
}

\maketitle

\begin{abstract}
 We study competition between firms in labor markets, following a combinatorial model suggested by Kelso and Crawford~\shortcite{KC82}. In this model, each firm is trying to recruit workers by offering a higher salary than its competitors, and its production function defines the utility generated from any actual set of recruited workers. We define two natural classes of production functions for firms, where the first one is based on additive capacities (weights), and the second on the influence of workers in a social network. We then analyze the existence of pure subgame perfect equilibrium (PSPE) in the labor market and its properties. While neither class holds the gross substitutes condition, we show that in both classes the existence of PSPE is guaranteed under certain restrictions, and in particular when there are only two competing firms. As a corollary, there exists a Walrasian equilibrium in a corresponding combinatorial auction, where bidders' valuation functions belong to these classes. 

While a PSPE may not exist when there are more than two firms, we perform an empirical study of equilibrium outcomes for the case of weight-based games with three firms, which extend our analytical results.
We then show that stability can in some cases be extended to coalitional stability, and study the distribution of profit between firms and their workers in weight-based games. 
\end{abstract}





%
%

\section{Introduction}

\emph{Labor markets} are matching games with two types of agents: firms and workers. Typically, each firm can recruit multiple workers, and is willing to compensate each worker by paying her salary for her work. An outcome in a labor market therefore consists of a partition of the workers between firms (possibly with some idle workers), and the salaries offered by the firms. 

The term \emph{worker} is quite general, and may refer to an employee, a consultant, a sub-contractor, or any other small service provider that the firm recruits. 
When considering interactions between a firm and its employees, the firm typically decides on the wage it offers to each employee. The employee, in turn, may either take the job for the proposed salary, or turn down the offer.

             
             Whereas most classical models of labor markets assume some uniform, divisible workforce (e.g.~\cite{JacSki04}), we will consider in this paper games with a finite number of firms and workers, following a model suggested by Kelso and Crawford~\shortcite{KC82}. This model allows for a rich combinatorial description of the synergies among workers working for the same company. In the most general case, an arbitrary value can be assigned to any group of workers and a particular firm. However, typically there is some structure to the value function (a.k.a. production function) that is derived from the context.
             
            As a simple example, consider a game where we can attribute some fixed ``productivity level'', capacity, or weight to each worker. 
             The value             generated by a firm is then some function of the total (additive)
            capacity of its workers. A concrete example for such a scenario
            is when firms compete for hiring suppliers that differ in ``weight'', which can be the amount of storage they supply, or number of patents they own. The utility of each firm is increasing in the total storage or total number of patents it acquires, regardless of how it is divided among the hired suppliers.

            
            
            Perhaps a
            more intricate scenario is when firms are recruiting influential
            members of a social network, trying to promote some new product. The
            value for a firm in this case is proportional to the joint influence
            of the recruited workers, which depends on the network structure and
            dynamics.

A natural question that arises in labor markets is the following: which outcomes, if at all, are stable? That is, which partitions and salaries guarantee that no firm or worker could gain by deviating from the current agreements. 
The answer of course depends on the game description (i.e. the value functions), but also on the bargaining dynamics and the allowed strategies. Kelso and Crawford~\shortcite{KC82} analyze a particular bargaining process, whose outcomes (in case of convergence) are referred to as the \emph{core}. They provide sufficient conditions for the existence of the core---namely, that all firms' value functions hold a technical property called \emph{gross substitutes} (See Appendix~\ref{sec:GS} for a formal definition). They also describe examples where it fails to exist. 

We consider a simple two-step dynamics, where at the first step firms commit to salaries, and in the second step workers choose which firm to join. These dynamics define a game in extensive form, which we call the \emph{competition game}. It turns out that the set of pure subgame perfect Nash equilibria (PSPE) in this two-step competition game coincides with the core, as defined in \cite{KC82}. Our primary goal is to characterize the conditions under which a PSPE exists for firms with value functions based on capacities or social connections. In particular, we are interested in extending the  results of Kelso and Crawford on existence of equilibrium to cases that are not covered by the gross-substitutes condition.

Nevertheless, we care not only about existence, but also ask what are the properties of such a stable outcome in terms of welfare, fairness, stability against group deviations, etc.

\subsection{Structure of the paper} 
In Section~\ref{sec:prelim} we formally define our two-step game, and classes of valuation functions inspired by the examples of competition over capacity and influence. We then introduce the equilibrium concepts used in the paper and explain in detail the connections between our model, the model of Kelso and Crawford~\shortcite{KC82} (henceforth, the ``KC model'') and other related models of competition. In particular, we show the equivalence between PSPE in our model, the core in the KC model, and Walrasian equilibrium in combinatorial auctions. 

In Section~\ref{sec:properties} we present some basic properties of PSPE in our setting and of the linear programs associated with it.

In Section~\ref{sec:weighted} we study games with additive capacities (weights) and show that an equilibrium always exists with two firms. A special case of weight-based games is uniform weights. This case, where all workers are homogeneous, entails gross-substitutes value functions, and therefore existence of PSPE with particularly natural properties follows from \cite{KC82}. We extend their result by showing that there always exists a \emph{coalition-proof} equilibrium in the homogenous case. 
In games with arbitrary capacities and more than two firms, a PSPE may not exist. However, the same natural strategy profile that yields an equilibrium in the previous cases can be generalized and used as a heuristic solution. We show through analysis and through an empirical study that this profile is usually quite stable, and that a PSPE almost always exists. We conclude the section with a study of the distribution of profit between firms and workers in equilibrium.
            
            In the social network model, studied in Section~\ref{sec:network}, we prove the existence of equilibrium for two firms when the network is sufficiently sparse.  We also demonstrate that our conditions for existence are minimal, in the sense that by relaxing any of them we can construct a game with no equilibrium.

Our results  are summarized in Table~\ref{tab:exist} on Page~\pageref{tab:exist}. We conclude in Section~\ref{sec:discussion}, where we discuss related work and the implications of our results, and suggest questions for future research.  In particular, we explain that our results entail the existence of Walrasian equilibrium in auctions where bidders have value functions based on either capacities or on influence in social networks. 

Due to space constraints and to allow continuous reading, some content has been deferred to the appendix. Omitted proofs are available in Appendix~\ref{sec:proofs}. Examples of games used in the paper can be found in Appendix~\ref{sec:examples}. The details of our experimental setting used for the emprical part of Section~\ref{sec:weighted}, appear in Appendix~\ref{sec:empirical}.

\section{Preliminaries}
\label{sec:prelim}
We denote vectors by bold lowercase letters, e.g.  $\vec a = (a_1,a_2,\ldots)$. Sets are typically denoted by capital letters, e.g. $B=\{1,2,\ldots\}$. When $\vec a$ is a vector of indexed elements and $B$ is a set of indices, we use the shorthand notation $a(B)=\sum_{b\in B}a_b$.

\emptypar In a \emph{competition game} we have 
a set $N$ of $n$ \emph{workers} and a set $K$ of $k$ \emph{firms}, where $k\geq2$. Each firm $i\in K$ is associated with a non-decreasing value function $v_i:2^N\rightarrow \mathbb R_+$, where $v_i(\emptyset)=0$. 
 The competition game  $G=\tup{N,K,(v_i)_{i\in K}}$ is a 2-step game in extensive form  as follows. 
\begin{description}
    \item[Step~1] Played by all firms simultaneously. The strategy of every firm $i\in K$ is a payment policy $\vec x_i$. That is,  $i$ promises to every worker $j$ that will join it a payment of $x_{ij}$.  
    \item[Step~2] Played by all workers simultaneously. The strategy of every worker $j\in N$ is choosing a firm $i\in K$, and is conditioned on the policies selected in step~1. Workers are allowed to remain idle by not choosing any firm.
\end{description}

Let  $\vec X\in \mathbb R^{k\times n}$ be the \emph{commitment matrix}, where $\vec X(i,j) = x_{ij}$.
\begin{itemize}
    \item A possible outcome of the game is $(P,\vec X)$, where $P=(S_0,S_1,\ldots,S_k)$ is a partition of workers to firms. $S_0$ contains workers that abstained (remained idle).
    \item Each worker $j$ that chose firm $i$, gains a utility of $x_{ij}$, where $x_{0,j} \equiv 0$. For each $j\in S_i$, we denote the realized payment to worker $j$ by $\hat x_j = x_{ij}$.
    \item Each firm $i$ gains a profit of $r_i(P,\vec X) = v_i(S_i) - \hat{x} (S_i)$.
\end{itemize}
Note that payments (promised or realized) may be negative in the general case. We denote by $SW(P)$ the sum of utilities of all the agents in a given partition. Note that for every $\vec X$, 
{\small
\begin{align*}
SW(P) &= SW(P,\vec X) = \sum_{i\in K}r_i + \sum_{j\in N}\hat x_j  \\
&= \sum_{i\in K}\(r_i + \sum_{j\in S_i}x_{ij}\) = \sum_{i\in K}v_i(S_i),
\end{align*}
}
i.e. the social welfare does not depend on $\vec x$, but only on the  partition of workers.

We restrict our attention to \emph{pure strategies} only. This includes both the payment policies of the firms, and the decisions of the workers.

\subsection{Value functions}
\label{sec:v_functions}
We use the notation $m_i$ for the marginal value of a worker to firm $i$. For every $S\subseteq N$ and $j\notin S$,
$m_i(j,S) = v_i(S\cup \{j\}) - v_i(S)$.

$v_i$ is \emph{submodular} (a.k.a. \emph{concave}), if $v_i(S \cup T)+v_i(S \cap T)\leq v_i(S)+v_i(T)$ for all $S,T\subseteq N$. It is \emph{strictly submodular} if the inequality is strict.
Equivalently, $v_i$ is submodular if the marginal contribution is nonincreasing. That is, for all $T\subseteq S$ and $j\notin S$,
$m_i(j,T)\geq m_i(j,S)$.  If the [strict] inequality holds only when $S,T$ are disjoint, then we say that $v_i$ is \emph{[strictly] subadditive}.
All value functions studied in this paper are submodular, i.e. have decreasing marginal returns. This assumption is standard in the economics literature~\cite{BR97,BMT02}. 

We say that workers $j$ and $j'$ are of the same \emph{type} if all firms are indifferent between them. That is, if for all $i\in K$, $S \subseteq N\setminus\{j,j'\}$, $v_i(S\cup \{j\}) = v_i(S\cup \{j'\})$. 
Similarly, we say that firms $i$ and $i'$ have the same type, if $v_i\equiv v_{i'}$. Games where all firms are of the same type are called \emph{symmetric} games.

\paragraph{Homogeneous games} In the simplest form of games, called \emph{homogeneous games}, all workers are of the same type. In such games each $v_i:[n]\rightarrow \mathbb R_+$ is a  function of the \emph{number} of the workers selecting firm $i$, i.e. $v_i(S) = v_i(|S|)$. 

\paragraph{Weighted games} The primary type of value functions we consider in this work is based on capacities, or \emph{weights}. 
Each worker has some predefined integer weight $w_j$, and the value of a set $S$ depends only on its total weight. Thus each $v_i$ is  a  function $v_i:\mathbb N\rightarrow \mathbb R_+$, where $v_i(S) = v_i\(\sum_{j\in S} w_j\)$. 

A game where all value functions are weight-based is called a \emph{weighted} game. Homogeneous games are a special case of weighted games, where all workers have the same weight (w.l.o.g. weight $1$).

A partition $P=(S_1,\ldots,S_k)$ of workers in a weighted game is \emph{balanced}, if the total weight of workers that choose each firm is the same, i.e. $w(S_i)=w(S_{i'})$ for all $i,i'$. A partition is \emph{almost balanced} if the total weight of any $S_i$ and $S_{i'}$ differ by at most $1$.

\paragraph{Influence in social networks}
Another value function we consider is inspired by social networks.
In a social network, we have a directed underlying graph, augmented with some deterministic or probabilistic process of information flow. The \emph{influence} of a set of nodes in the graph is the expected number of nodes that get a message broadcasted by the original set (assuming some specific dynamics, for example one of the models suggested by Kempe et al.~\shortcite{KKT03}).  Every social network then induces a competition game, where the workers are some particular set of influential nodes for hire, and firms try to recruit workers with maximal influence in the network. For the formal model see Section~\ref{sec:network}.
%

\paragraph{Synergy graphs}
Simple synergies between workers (or between items in a combinatorial auction) can be represented by a weighted undirected graph, where every vertex corresponds to a worker, and the value of a set is the total weight of edges  linked to vertices in the set. This includes edges between vertices in the set, and edges between these vertices and external vertices.\footnote{If we only consider edges within the set, the value function coincides with that of \emph{induced subgraph games}, as defined by Deng and Papadimitriou~\shortcite{DengPa94}. While our definition of synergy graphs implies a submodular value function, the value in induced subgraph games is supermodular.} In Section~\ref{sec:network} we show that competition games induced by synergy graphs coincide with a particular type of social network games. 

All classes described above---except homogenous valuations---may violate the gross-substitute condition. See Example~\ref{ex:weighted_no_GS} in the appendix for details.

\subsection{Equilibrium concepts}
\label{sec:refine}
In a general game of the form we described there can be many pure Nash equilibria, but most of them are uninteresting. Formally, workers are allowed to play any strategy of the 2-step game (i.e. commit to some partition for every possible choice of firms' policies). Then the following, for example is a Nash equilibrium: firms keep all profit to themselves, while workers play an arbitrary fixed partition. In the other extreme, workers can commit to go only with firms that keep at most  $\eps$ for an arbitrarily small $\eps>0$ (if there is more than one such firm, all workers pick the firm with the smallest index). If all firms charge more, all workers go to the one that charges the least. In equilibrium, firm~1 gets all workers, but keeps a profit of $\eps$ from each.

\paragraph{Subgame-perfect Nash equilibrium}
It is clear from the observation above that many Nash equilibria are not really plausible, especially when there are no binding agreements between workers. We therefore need some equilibrium refinement, and the natural candidate is \emph{pure subgame perfect Nash equilibrium} (PSPE). Indeed, this means that the workers are not allowed to make non-credible threats, and for any choice of policies must play an equilibrium strategy (of $N$), given these policies. 

Formally, given any payment matrix $\vec X$ 
we denote by $\cal P(\vec X)$ the set of partitions where each worker plays a weakly dominant strategy w.r.t. $\vec X$. That is, each $j\!\in\! N$ selects a firm  $i$ offering the highest payoff. We assume workers use a known tie-breaking rule, so that the result of offering payoffs of $\vec X$ is a unique well-defined partition $P(\vec X)\in \cal P(\vec X)$. 

\begin{definition}
\label{def:PSPE}
An outcome $(P,\vec X)$ is a \emph{pure subgame perfect equilibrium} (PSPE) if (a) $P=P(\vec X)$; and (b) for every $i\in K$ and $\vec x'_i$, $r_i(P',\vec X') \leq r_i(P,\vec X)$, where $\vec X' = (\vec x_{-i},\vec x_i), P'=P(\vec X')$.
\end{definition}

Since the behavior of the workers is straight-forward (playing according to $P(\vec X)$, the game effectively reduces to the first step. 

\begin{observation}
\label{ob:max}
Let $G$ be a competition game, and let $(P,\vec X)$ be a PSPE. Then for all $i\in K$ and $j\in S_i$, there is at least one additional firm $i'\neq i$ (and w.l.o.g. all firms), that is offering $x_{i'j} = x_{ji} =\hat x_j$ (otherwise $i$ could have offered a lower amount). The PSPE outcome $(P,\vec X)$ can therefore be written as $(P,\vec x)$, where $x_j = \hat x_j$ for all $j\in N$. This observation greatly facilitates the analysis of  PSPEs.
\end{observation}

\paragraph{The core}
Kelso and Crawford~\shortcite{KC82} define the core of a labor market as an outcome (partition of workers and salaries) from which there is no group that can deviate in a way that makes all of them strictly gain. In the model they consider,\footnote{In the KC model workers may have idiosyncratic preferences over firms. Our model is a special case where all firms are treated equally by workers. } they show that w.l.o.g. it is sufficient to consider groups of one firm and several workers.

\begin{lemma}
\label{lemma:PSPE_core}
An outcome $(P,\vec x)$ is a PSPE in our two-step game if and only if it belongs to the core of the corresponding labor market in the KC model. 
\end{lemma}
\begin{proof}
Indeed, if there is a firm $i\in K$ and workers $C\subseteq N$ that can gain by deviating in the KC model using payoffs $(x'_{ij})_{j\in C}$, then firm $i$ can offer salary $x'_{ij}>\hat x_{j}$  in the two-step game to each $j\in C$ (and $0$ to every other worker). In the second step all workers of $C$ will move to $i$ according to their subgame-perfect strategy.\footnote{Note that in our model this is considered a deviation of a single player---the firm $i$.} In the other direction, if some firm $i\in K$ can deviate from $(P,\vec x)$ in the two-step game, and remains with a set of workers $S'_i$, then the coalition $\{i\} \cup S'_i$ violates the core constraints. 
\end{proof}

\begin{table*}
\begin{tabular}{lccc}
\cline{2-2}
\cline{4-4}
Labor  markets:  &         \multicolumn{1}{|c|}{Our model}  & (a) & \multicolumn{1}{|c|}{ the KC model}  \\
      (firms, workers)      &         \multicolumn{1}{|c|}{   \emph{PSPE} }      &  $\leftrightarrow$ &  \multicolumn{1}{|c|}{ the \emph{core}} \\
\cline{2-2}
\cline{4-4}
				    &   $\updownarrow$ (d) &			~~~~~~~~~~~~~~~~~~		 &    $\updownarrow$ (b) \\
\cline{2-2}
\cline{4-4}
Combinatorial  auctions:   &    \multicolumn{1}{|c|}{ First price auction with sealed bids} &    (c)    &    \multicolumn{1}{|c|}{  the Arrow-Debreu model} \\
     (bidders, items)               &       \multicolumn{1}{|c|}{ \emph{pure Nash equilibrium} }   &    $\leftrightarrow$ & \multicolumn{1}{|c|}{ \emph{Walrasian equilibrium}}\\
\cline{2-2}
\cline{4-4}
\end{tabular}
\caption{\label{tab:eq_concepts}
The relations between the various models, and their respective equilibrium concepts (in \emph{italics}). The equivalences follow from: (a) Lemma.~\ref{lemma:PSPE_core}; (b) \cite{GulS99}; (c)  \cite{Has+11}  ; (d)  Obv.~\ref{ob:PSPE_auction}.}
\end{table*}

\paragraph{Combinatorial auctions}
One can think of an alternative interpretation of our model, where firms are bidders in a combinatorial auction, and the workers are the ``items''. The valuation of a bundle $S$ of items to each bidder $i$ is defined by $v_i(S)$.  Every bidder submits a sealed bid for each item, which goes to the highest bidder.
\begin{observation}\label{ob:PSPE_auction}
Our two-step game between firms and workers is equivalent to a first price auction over the items (a separate but simultaneous auction for each item). Further, every PSPE coincides exactly with a Nash equilibrium of the auction. 
\end{observation}
Indeed, the dominant strategy of each worker in the second step is equivalent to the auction rule that every item must go to the highest bidder. A deviation of a firm in the first step is equivalent to a deviation of a bidder to an alternative bidding vector. 
Further, it is known that the Nash equilibrium of the first price combinatorial auction (and hence also our PSPE) coincides with a Walrasian equilibrium in the Arrow-Debreu model, in terms of item (worker) allocation and the realized prices $(\hat x_1,\ldots,\hat x_n)$; and that Walrasian equilibrium coincides with the core of the KC model. See Table~\ref{tab:eq_concepts} for an illustration of the connections between the models.

By this correspondence between the core, Walrasian equilibrium and PSPE, existence results translate directly from one model to the other. 

\paragraph{Coalition-proof equilibrium}
Standard PSPE, just like Nash equilibrium, guarantee that no \emph{single} agent can gain by deviating from the outcome. However, it does not preclude the formation of \emph{coalitions}, who will make a joint deviation in an attempt to bias the outcome. \emph{In contrast with the core} of the KC model, coalitions of two or more firms in our model may have substantially more power. A coalition of all firms can decide, for example, to cut all salaries by half without the approval of the workers.

 An intrinsic problem with the formation of such coalition, is that the coalition itself may not be stable. That is,  it may contain an agent (or a subset of agents) that will change its strategy again, knowing how the other members of the coalition will play. A solution concept that was introduced in particular to demonstrate resistance to such unstable collusion is \emph{coalition-proof equilibrium}~\cite{CProof87}. We slightly modify the definition to fit in our context.

\begin{definition} \label{def:CP_PSPE}
An outcome of competition game $(P,\vec X)$ is a \emph{Coalition-Proof subgame perfect equilibrium} (CP-PSPE), if 
(a) The outcome $(P,\vec X)$ is a PSPE; and 

\noindent (b) Consider a coalition $R=K' \cup N'$ and a deviation of $R$ to a (pure) profile $\vec z_{K'}$, $Q_{N'}$ that induces the outcome $(Q,\vec Z)= ((P'_{-N'},Q_{N'}),(\vec x_{-{K'}}, \vec z_{K}))$ where $P'=P(\vec Z)$ (i.e. non-deviating workers select each the highest paying firm). If all of $R$ strictly gain by this deviation, then there is some agent $g\in R$ (either a firm or a worker) that can  strictly gain by deviating from $(Q,\vec Z)$ to some other strategy.
\end{definition}
Our definition is stronger (i.e., implies a higher level of stability) than the standard definition of coalition-proofness in the sense that the profile must also be a PSPE, and also since it only allows a single agent to re-deviate from the coalition $R$.

Finally, we define the outcome $(P,\vec X)$ as a \emph{cartel-proof PSPE}, if it is coalition-proof  only w.r.t. coalitions of firms (without workers). Thus cartel-proofness is somewhat weaker than coalition-proofness.

\begin{lemma}
\label{lemma:cartel_proof}
$G$ has a PSPE if and only if $G$ has a cartel-proof PSPE.
\end{lemma}

We emphasize that cartel-proof PSPEs do not necessarily coincide with the KC core. For example, firms may deviate together by lowering all  salaries by a small constant. 

\paragraph{Policy types}
We also consider other natural requirements that policies may hold. 
 A payment policy $\vec x_i=(x_{i,1},\ldots,x_{i,n})$ is \emph{fair} if for every  pair $j,j'$ of the same type, it holds that $x_{ij} = x_{ij'}$. Fair policies are necessary for example in some Internet settings, where the firm can verify a worker's type, but not her identity.
 In \emph{weighted games}, a policy $\vec x_i$ is \emph{proportional}, if for all $j,j'$ it holds that $\frac{x_{ij}}{x_{ij'}} = \frac{w_j}{w_{j'}}$. Any proportional policy is fair.

We say that an  outcome $(P,\vec X)$ is \emph{fair} [respectively: \emph{proportional}], if all the policies $\vec x_1,\ldots, \vec x_k$ are fair [resp.: proportional]. 
Note that we do not externally enforce fairness nor proportionality.


\section{Properties of equilibrium outcomes}
\label{sec:properties}
PSPEs have many desired properties, which motivate the search for such outcomes. In addition, some of these properties will be used as tools in the next sections to prove existence and non-existence of  PSPEs in various games.


The first property is a technical condition, which basically means that a firm cannot gain by recruiting or releasing a single worker. 
\begin{lemma}
\label{lemma:marginal}
Let $(P,\vec x)$ be a PSPE outcome in game $G$. Then for all $i\in K$ and $j\in S_i$,
\begin{enumerate}
    \item $ x_j \leq m_i(j,S_i\setminus \{j\})$.
    \item For any $i'\neq i$, $ x_j \geq m_{i'}(j,S_{i'})$.
\end{enumerate}
\end{lemma}

%

\subsection{Individual rationality, Fairness and Envy freeness}
\label{sec:EF}
We next present three simple observations (phrased as lemmas), showing that a PSPE outcome is always individually rational, envy free, and that w.l.o.g. it is fair.

Workers always have the option to remain without a firm and get a payment of $0$. This means that a worker will never join (in equilibrium) a firm that is offering a negative payment. Firms can preclude certain workers from joining  their coalition, by offering them a strictly negative payment.

As a result, the profit of a firm in equilibrium is also non-negative, as it can always do better by refusing all workers. The following observation concludes that no agent is harmed by participating in the competition game.

\begin{lemma}[Individual rationality]
\label{lemma:IR}
Let $(P,\vec x)$ be a PSPE outcome in game $G$, then (1) $x_j\geq 0$ for all $j\in N$; and (2) $v(S_i)-x(S_i)\geq 0$ for all $i\in K$.
\end{lemma}

We say that firm~$i$ \emph{envies} firm~$t$ in an outcome $(P,\vec x)$, if $i$ wants to trade workers and payments. That is, if
$v_i(S_{t})-\sum_{j\in S_t}x_{j} > v_i(S_i)-\sum_{j\in S_i}x_{j} (=r_i(P,\vec x)).$
An outcome is \emph{envy-free} if no firm envies any other firm.

\begin{lemma}[Envy freeness]
\label{lemma:EF}
Let $(P,\vec x)$ be a PSPE outcome in game $G$. Then $(P,\vec x)$ is envy-free.
\end{lemma}
The above holds because an envious firm $i$ can always dismiss its workers $S_i$ and recruit $S_t$ instead. Since payments are fixed, $i$ only needs to pay more than $\sum_{j\in S_t}x_j$ and with a sufficiently small extra payment this is a profitable deviation. The proof of Lemma~\ref{lemma:fair} appears  in the appendix.

\begin{lemma}[Fairness]
\label{lemma:fair}
If $(P,\vec x)$ is a PSPE in game $G$, then there is a \emph{fair} outcome $(P,\vec x^*)$ that is a PSPE in $G$, where the profit of each firm remains the same.
\end{lemma}

\subsection{LP formulation}
\label{sec:LP}
There are two natural ways to describe a PSPE with a linear program.

\paragraph{ILP relaxation and the welfare theorems}
Computational schemes for representing combinatorial markets and to solve them (i.e., to compute a Walrasian equilibrium) have been thoroughly studied (see Blumrosen and Nisan~\shortcite{BlumNisan07} for details).
Since PSPE coincides with the Walrasian equilibrium of a corresponding combinatorial market (as explained in Section~\ref{sec:refine}), we can infer some important properties of PSPEs, and apply standard techniques from the literature for computing them. The first and most fundamental property is about efficiency.

\begin{atheorem}{First welfare theorem (FWT)}
Every Walrasian equilibrium, if exists, is optimal in terms of the social welfare.
\end{atheorem}

As an immediate corollary from the first welfare theorem, we get that if $(P,\vec x)$ is a PSPE, then $P$ necessarily maximizes the social welfare.

There is a standard Integer Linear Program, denoted $ILP(G)$, whose solutions describe the optimal partition in the game. 
The \emph{Linear Program Relaxation} of $ILP(G)$ is denoted by $LPR(G)$. The \emph{second welfare theorem} (SWT) states that a Walrasian equilibrium (and  thus PSPE) exists if and only if the intergrality gap of $ILP(G)$ is zero, i.e. if the solution quality of $ILP(G)$ and $LPR(G)$ is the same. Moreover, in such cases it is known that the solutions to the dual linear program of $LPR(G)$ yield the market clearing prices, which correspond to the equilibrium strategies (salaries) of the firms in PSPE.

From a computational complexity perspective, it is known that if demand queries for every $v_i$ can be computed in polynomial time,\footnote{ The input of a demand query over a valuation function $v_i$ is a payoff vector $\vec x$, and its output is the optimal set of workers for these payoffs, i.e. $\argmax_{S\subseteq N}v_i(S)-x(S)$.} then $LPR(G)$ (and thus also $ILP(G)$ when an equilibrium exists) can be solved efficiently. 

\paragraph{Partition dependent program}
While solving $LPR(G)$ is an elegant way to handle the problem, it may be quite challenging in practice, especially when demand queries are hard to compute. We propose a simple linear program that is designed to find the equilibrium payoffs, given an optimal partition. If an equilibrium fails to exist, or if the given partition is not optimal, the program will have no valid solutions.

For every potential deviation of a firm (i.e. dismissing several workers and recruiting others), we can write the requirement that the firm does not gain as a linear constraint. We denote the linear program corresponding to a partition $P$ in game $G$ by $LP(G,P)$, over the variables $x_1,\ldots,x_n$. $LP(G,P)$ has the following constraints:
$$\forall i\in K,\forall A\subseteq S_i,\forall B\subseteq N\setminus S_i,~~ v_i(S_i)- x(A) \geq v_i((S_i\setminus A)\cup B) - x(B).$$
The constraints in Lemmas~\ref{lemma:marginal} and \ref{lemma:IR} are special cases. This linear program can be used to compute equilibrium outcomes in a given game once an optimal partition is known, as detailed in Section~\ref{sec:compute}. It has the additional advantage over $LPR(G)$, that we can tailor the optimization goal to our needs. For example, to maximize firms' revenue or to minimize it.

\subsection{Computation and complexity}
\label{sec:compute}
Suppose we want to compute a PSPE for a given competition game $G$. By SWT, the standard way of computing an equilibrium when one exists, is to solve $LPR(G)$ (or, more precisely, its dual). However, there are several drawbacks to this approach. First, we do not have an explicit representation of the partition, and we do not know if it is a legal solution to the original problem $ILP(G)$ (i.e. if a PSPE exists). Second, even in cases where we know that a PSPE must exist, efficiently solving $LPR(G)$ requires an efficient algorithm to compute demand queries on $v_i$. Unfortunately, computing the optimal partition---and thus solving a demand query---is NP-hard even for weighted games. To see this, observe that finding the optimal partition in a weighted game with two symmetric firms coincides with the {\sc Partition} problem.\footnote{Similarly, computing the optimal partition for two firms in a symmetric game over a synergy graph is equivalent to the NP-hard problem of {\sc MaxCut}.} Finally, there may be more than one PSPE and we want to study the full range of equilibria.

We therefore apply a simpler computation method. By FWT, we can divide this task into two parts: first find an optimal partition $P$, and then compute a stable payment vector $\vec x$ by solving the linear program $LP(G,P)$. While both tasks are still computationally challenging in the general case, we can make some simplifying assumptions. If we can limit the number of firms and/or the number of workers' types, then efficient algorithms exist. See Proposition~\ref{th:compute_types} in the appendix for details. 



%
In weighted games we can further exploit symmetries among workers for a more efficient computation (see Proposition~\ref{th:weighted_compute}).

\section{Weighted games}
\label{sec:weighted}
The first class of value functions we study is based on capacities, or weights.
Recall that a weight based value function $v:\mathbb N \rightarrow \mathbb R_+$ is a subadditive functional, which maps the total weight of a set of workers $w(S)$ to utility (thus $v(S)$ is submodular). We sometimes write $v$ as a vector of $w(N)+1$ entries $(v(0),v(1),\ldots,v(w(N)))$, where by convention, the first entry $v(0)=0$.

Before continuing to our existence results, we observe that without the subadditivity assumptions, weighted games may not posses a PSPE even in a most simple scenario. Indeed, consider a homogeneous and symmetric game with a weighted (non-subadditive) value function $v= (0,3,4,6)$,  two firms, and three workers. It is easy to verify using the properties showed in the previous sections that this game has no PSPE (see Example~\ref{ex:nonsub} in the appendix for details).

\paragraph{Homogeneous games}
Suppose that all workers have unit weight. Kelso and Crawford~\shortcite{KC82} show that in such games the core is always non-empty. It follows that a PSPE always exists. Moreover, when firms are symmetric, then such a PSPE has a particularly simple form. 

Indeed, it is not hard to see that the most efficient partition of workers (in terms of social welfare) is one that is almost balanced. Otherwise, a worker can improve the welfare by moving from an overloaded firm to one that has two workers less. Let $q=\floor{n/k}$ and $\delta=v(q+1)-v(q)$. By FWT, in any PSPE every firm has either $q$ or $q+1$ workers. Also, by Lemma~\ref{lemma:marginal}, there is a PSPE where the payoff to every worker is $\delta$ (and unless the partition is exactly balanced, this is the unique PSPE).



\subsection{The case of two firms}
\label{sec:weighted_two}


Consider a symmetric weighted game with two firms and only two workers $N=\{h,l\}$, where $w_h \geq w_l$.
This simple case can be solved as follows. Let $x_l = v(w_l+w_h)-v(w_h)$, and $x_h = v(w_l+w_h)-v(w_l)$, i.e. we set the payment of each worker to be its own marginal contribution to the set $N$. Then, the policy of each firm is to pay $x_j$ to each $j\in N$. It is easy to verify that these policies admit two PSPEs, where in each there is one worker per firm.

However, the described policy is not necessarily proportional. A proportional outcome, which is also a PSPE, would be to pay $x_j=v(w_j)$ to each worker $j\in \{l,h\}$. If $v$ is strictly subadditive, then there are also other proportional PSPEs where the firms keep some of the profit. We next show that we can always find a proportional PSPE for two firms and any number of workers. Note that we only require that each value function will be subadditive.

We next lay our main existence result for the weighted setting.
\begin{theorem}
\label{th:PSPE_fair_k2_distinct}
Let $G=\tup{N,K,\vec w,v_1,v_2}$ be a weighted competition game with two firms, then $G$ admits a proportional PSPE.
\end{theorem}
 The proof hinges on the idea of computing the marginal value of a \emph{unit of weight}. However in the general case this is an evasive notion that requires a nontrivial case analysis (see Appendix~\ref{sec:proofs_weighted}). We bring here a simplified proof of the symmetric case.
\begin{proof} [sketch of the symmetric case]
Indeed, let $P=(S_1,S_2)$ be an optimal partition.  Denote $H= w(S_1)$, $L = w(S_2)$, and suppose that $L\leq H$. By optimality, the gap $H-L$ is minimal. We define $\delta = \frac{v(H)-v(L)}{H-L}$, and argue that it induces a PSPE $(P,\vec x)$, where $\vec x = \delta\cdot \vec w$. Indeed, the profit of a firm with workers of total weight $q$ is $r(q) = v(q)-\delta q$. This is a concave function in $q$, with maximum in $q^*= \floor{W/2}$. Denote $\Delta=H-L$. In $P$ we have 
{\small
\begin{align*}
\Delta r_1 &= \Delta(v(H) - H\delta) = \Delta v(H) - H (v(H) - v(L)) \\
&= H\cdot v(L) - L\cdot v(H)&\\
\Delta r_2 &= \Delta (v(L) - L\delta) = \Delta v(L) - L (v(H) - v(L)) \\
&=  H\cdot v(L) - L\cdot v(H) = \Delta r_1 
\end{align*}
}
This means that $r_1=r_2$, and by minimality of the gap, firms cannot get closer to the theoretical optimal profit $r(q^*)$: if a firm $i$ deviates, it necessarily ends up with a set of workers $S'_i$, s.t. $|w(S'_i)-q^*| \geq |w(S_i) - q^*|$, and thus $r'_i = r(q'_i) \leq r(q_i) = r_i$.
\end{proof}


\subsection{More than two firms}
\label{sec:weighted_k}

A question that naturally arises is whether we can generalize Theorem~\ref{th:PSPE_fair_k2_distinct}, i.e.  prove that a PSPE always exists for any number of firms, perhaps even with the additional requirement of proportionality. 
A  result by Gul and Staccheti~\shortcite{GulS99} shows that whenever there is a firm/bidder whose value function violates gross-substitutes, it is possible to construct an example (with additional unit-demand bidders) where an equilibrium does not exist. While their construction does not apply directly to our case, it gives little hope that a PSPE exists in the general case of weighted games.

Indeed, Proposition~\ref{th:no_PSPE_k3} below shows that a PSPE is not guaranteed for multiple firms. 
We first put forward a simple example showing that a \emph{proportional} PSPE may not exist.  
\begin{example}
\label{ex:prop_k3}
Consider a symmetric game $G^*$ where $w_1=5,w_2=6,w_3=7$, and $v(5)=124,v(6)=126,v(7)=127$. Clearly in the optimal partition each firm has a single worker. Suppose there is some proportional PSPE, where $x_j = \delta\cdot w_j$ for all $j$. Then by Envy-freeness (Lemma~\ref{lemma:EF}), all three firms make the same profit, i.e. 
$$r_1 = r_2 = r_3 \Rightarrow v(5)-5\delta \stackrel{(\#1)}{=} v(6)-6\delta \stackrel{(\#2)}{=} v(7)-7\delta.$$
By  Eq.~($\#1$), $\delta = v(6)-v(5) = 2$, whereas by ($\#2$), $\delta = v(7)-v(6) = 1$. A contradiction.
\end{example}

\begin{proposition}
\label{th:no_PSPE_k3}
For any $k\geq 3$, there is a symmetric weighted competition game with $k$ firms that does not have a  PSPE at all.
\end{proposition}
An example proving the proposition appears in the appendix (Example~\ref{ex:sym_3_no_PSPE}). We hereby construct an example with $k=4$ firms. 

\begin{example}\label{ex:sym_4_no_PSPE}
Our game $G$ has $9$ workers in total, where the weights are $(2,2,2,2,2,3,3,3,5)$. We define $v(w)=\min\{w,6\}$.\footnote{In fact, most subadditive functions will work. We selected one that simplifies the argument.} We claim that the optimal partition must be either $P_1 = \(\{5\},\{3,2,2\},\{3,3\},\{2,2,2\}\)$ or $P_2 = \(\{3,2\},\{5,2\},\{3,3\},\{2,2,2\}\)$ (up to permutations of agents of the same type). See Fig.~\ref{fig:k4_example}.

By the shape of $v$, the optimal partition minimizes $|\{i\in K : w(S_i) < w'\}|$ for $w'=1$, then for $w'=2,3,4$, etc. Indeed, the total weight is $24$, but it cannot be divided in a balanced way. 
Thus we get that every optimal partition has total weights of $(5,7,6,6)$. $P_1$ and $P_2$ are the only ways to construct such a partition. In terms of the integrality gap, this is an example of a weighted game with $IG(G)=\frac{23.5}{23}=\frac{47}{46}$. 
Next, we want to assign payments. 
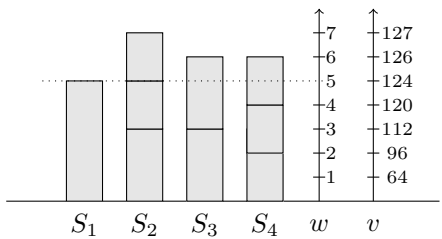
\begin{figure}[h]
\begin{center}


\begin{tikzpicture}[scale=1.6]

\tikzstyle{dot}=[rectangle,draw=black,fill=white,inner sep=0pt,minimum size=4mm]

%
%


%
\draw (-2,0) -- (1.6,0);
\draw[fill=black!10!white] (-1.5,0) rectangle (-1.2,1);
\node at (-1.35,-0.2)  {$S_1$};

\draw[fill=black!10!white] (-1,0) rectangle (-0.7,0.6);
\draw[fill=black!10!white] (-1,0.6) rectangle (-0.7,1);
\draw[fill=black!10!white] (-1,1) rectangle (-0.7,1.4);
\node at (-0.85,-0.2)  {$S_2$};

\draw[fill=black!10!white] (-0.5,0) rectangle (-0.2,0.6);
\draw[fill=black!10!white] (-0.5,0.6) rectangle (-0.2,1.2);
\node at (-0.35,-0.2)  {$S_3$};

\draw[fill=black!10!white] (0,0) rectangle (0.3,0.6);
\draw[fill=black!10!white] (0,0.4) rectangle (0.3,0.8);
\draw[fill=black!10!white] (0,0.8) rectangle (0.3,1.2);
\node at (0.15,-0.2)  {$S_4$};

\draw[->] (0.6,0) -- (0.6,1.6);
\node at (0.6,-0.2) {$w$};
\node at (0.7,0.2) {\scriptsize{$1$}};
\draw (0.55,0.2) -- (0.65,0.2);
\node at (0.7,0.4) {\scriptsize{$2$}};
\draw (0.55,0.4) -- (0.65,0.4);
\node at (0.7,0.6) {\scriptsize{$3$}};
\draw (0.55,0.6) -- (0.65,0.6);
\node at (0.7,0.8) {\scriptsize{$4$}};
\draw (0.55,0.8) -- (0.65,0.8);
\node at (0.7,1) {\scriptsize{$5$}};
\draw[dotted,thin] (-1.7,1) -- (0.65,1);
\node at (0.7,1.2) {\scriptsize{$6$}};
\draw (0.55,1.2) -- (0.65,1.2);
\node at (0.7,1.4) {\scriptsize{$7$}};
\draw (0.55,1.4) -- (0.65,1.4);

\draw[->] (1.05,0) -- (1.05,1.6);
\node at (1.05,-0.2) {$v$};
\node at (1.25,0.2) {\scriptsize{$64$}};
\draw (1.0,0.2) -- (1.1,0.2);
\node at (1.25,0.4) {\scriptsize{$96$}};
\draw (1.0,0.4) -- (1.1,0.4);
\node at (1.25,0.6) {\scriptsize{$112$}};
\draw (1.0,0.6) -- (1.1,0.6);
\node at (1.25,0.8) {\scriptsize{$120$}};
\draw (1.0,0.8) -- (1.1,0.8);
\node at (1.25,1) {\scriptsize{$124$}};
\draw (1.0,1) -- (1.1,1);
\node at (1.25,1.2) {\scriptsize{$126$}};
\draw (1.0,1.2) -- (1.1,1.2);
\node at (1.25,1.4) {\scriptsize{$127$}};
\draw (1.0,1.4) -- (1.1,1.4);


%
%
%
%
%

\end{tikzpicture}
\caption{\label{fig:k4_example} The optimal partition $P_1$.}
\end{center}
\end{figure}

By Lemma~\ref{lemma:fair}, we can set a uniform payment to each type of workers, thus we should determine the values of $x_2,x_3$ and $x_5$.
 Now, by Envy-freeness (Lemma~\ref{lemma:EF}), $r_3=r_4$, and thus
$0 = r_3-r_4 = (v(6)-3x_2) - (v(6)-2x_3) = 3x_2-2x_3.$
Similarly, firms~1 and 2 can trade the 5-worker for the set $\{2,3\}$. Thus $x_5=x_2+x_3$.
 We get that $\vec x$ must be a proportional payment vector, where $\vec x = \delta \vec w$ for some unit payoff $\delta$. However, firms~1,2 and 3 have a total weight of $5$,$7$ and $6$ respectively, exactly as in the game $G^*$ in Example~\ref{ex:prop_k3}. As we show above, such a proportional payment vector cannot be stable.\hspace*{\fill} $\diamond$
\end{example}

The last result seems discouraging, as in certain situations a stable outcome may not exist at all. Nevertheless, this is a worst-case outcome. We argue that such cases are quite rare and do not represent the common situation.  We assert that at least for the case of symmetric firms, a  PSPE usually exists. We support this assertion using both a formal argument and an experimental study. First, we prove that if the weights of the workers allow for the formation of an \emph{almost balanced partition}, then we can find a proportional PSPE, regardless of the value function $v$.  Note that when weights are small integers,  there is typically an almost-balanced partition.\footnote{For example, if there are at least $k\cdot \max_j w_j$ workers with weight $1$, then an almost-balanced partition \emph{must} exist.}
We then generate weighted games from a simple distribution, showing that they usually admit a  PSPE.

\begin{proposition}
\label{th:PSPE_balanced}
Let $G=\tup{N,K,\vec w,v}$ be a weighted competition game with $k$ symmetric firms. (a) If there is an almost-balanced partition of $\vec w$, then $G$ admits a proportional PSPE. (b) symmetry is a necessary condition.
\end{proposition} 
The proof of part (a) is in Appendix~\ref{sec:proofs_weighted}.
Part (b) is demonstrated by Example~\ref{ex:asym_balanced} in the appendix, which shows that a PSPE may not exist in asymmetric games, even when there is a fully balanced partition.
%
%
%
%
%

\paragraph{Experimental study of stability with three firms}
The full details of our experimental setting appear in Appendix~\ref{sec:emp_setting}, and our results on stability are in Appendix~\ref{sec:emp_heuristic}. We only bring here a short summary of the results.

We generated over 6000 random instances of weighted competition games with three symmetric firms and up to 14 workers. Although less than a third of the instances admitted an almost balanced partition, in all but 4 instances  a PSPE was found. 

In fact, in most generated instances,  there is a proportional PSPE that can be found using simple heuristics. Recall that for two symmetric firms, we computed the ``unit payoff'' $\delta$, as the average marginal contribution of a single unit of weight in the optimal partition. That is, let $P^*$ be the optimal partition, $i_+ = \argmax_{i\in K} w(S_{i})$, $i_- = \argmin_{i\in K} w(S_{i})$, and $d=w(S_{i_+})-w(S_{i_-})$. Then $\delta=\frac{v(w(S_{i_+})) - v(w(S_-))}{d}$.  In the induced proportional payoff vector, $x^*_j = \delta\cdot  w_j$ for all $j\in N$. 

Note that for any weighted game we can compute $(P^*,\vec x^*)$ and use it as a \emph{heuristic} solution, hoping that no firm will have a strong incentive to deviate.
By Proposition~\ref{th:PSPE_balanced}, this heuristic solution is always a PSPE when the gap $d$ is at most $1$ (i.e. $P^*$ is almost-balanced). The experimental results show how its performance gradually deteriorates as $d$ is increasing. 

It turns out that in more than half of the total tested instances (including third of the non-almost-balanced instances), the maximal gain is 0. That is, $(P^*,\vec x^*)$ is a  PSPE. Further, only in 6\% of the generated instances there was an agent (firm) that could improve its profit by more than 5\%. In other words, $(P^*,\vec x^*)$ is an $\eps$-PSPE outcome for $\eps<0.05$ for over 94\% of the instances (when the profit $r$ is normalized to $1$).

We note that the size of the gap $d$ tends to decrease as we increase the diversity of workers (i.e. the number of types). Therefore, if we avoid limiting the number of types then the heuristic solution works even better. 

We do not claim that our random sample is characteristic of every possible scenario. It may well be the case that with more than three firms, or with a different distribution of workers' weights, a higher number of ``bad'' instances emerge. We do feel however that these experimental results strengthen the conclusion, that when faced with a given weighted game, then (a) it is very likely to have a  PSPE; and (b)  heuristic proportional payoffs will usually be quite stable.


\subsection{Coalitional stability}

\paragraph{Homogenous games}
As mentioned above, existence of PSPE in this case follows as a special case from the existence of the core in the KC model. However, in our model this result can be significantly strengthened as follows. 





Since a PSPE exists, we know by Lemma~\ref{lemma:fair} that a \emph{fair} PSPE also exists. It follows that there is some value $\delta=\delta(G)$ and a partition $P=P(G)$, s.t. if workers split according to $P$ and are paid $x_j=\delta$ each, no worker or firm wants to deviate. For a given game $G$, let $(P^*,\boldsymbol \delta^*)$ be the PSPE of this form, where $\boldsymbol \delta^* = \boldsymbol \delta^*(G) = (\delta^*,\delta^*,\ldots,\delta^*)$ for the lowest possible value of $\delta^*$.

\begin{proposition}
 Let $G$ be a homogeneous competition game. Then $(P^*,\boldsymbol \delta^*)$ is a  CP-PSPE in $G$.
\label{th:CP_PSPE}
\end{proposition}
\begin{proof} 
W.l.o.g. all workers weigh $1$.
Assume, toward a contradiction, that there is a coalition $R=K'\cup N'$ that can deviate from the proposed solution to some other partition $P'=(S'_1,\ldots,S'_k)$ and a payoff vector $\vec x'$. 
We now divide into several cases.

\subpar{Case~1} Suppose first that $K'\neq K$, i.e. at least one firm is not part of the coalition $R$. 
Consider any $i\in K'$. Since there is at least one firm that keeps the original strategy, $i$ must pay at least $\delta^*$ to all of its workers $j\in S'_i$, and strictly more than $\delta^*$ if $j\in N'$ (regardless of the strategies of $K'\setminus \{i\}$). Therefore $i$ cannot strictly gain, as this would entail a unilateral deviation of $i$ from the PSPE profile $(P^*,\vec \delta^*)$.

\subpar{Case~2}  $K\subseteq R$.
Denote by $A_i\subseteq S'_i$ all workers that get strictly less than $\delta^*$, i.e. $\hat x'_j < \delta^*$. We claim that $A_i\neq \emptyset$ for all $i\in K$. Indeed, if a firm pays at least $\delta^*$ to each worker, it cannot improve its profit (once again using the above argument). Let $\eps>0$ s.t.  $\hat x'_j \leq \delta^*-\eps$ for all $j\in A_i$. 
 Note that since workers in $A_i$ are worse off in the new outcome, they cannot be deviators. Thus $A_i$ maintain their original strategy (to prefer the firm offering the highest payment).
 
\subpar{Case~2a} $P'\neq P^*$. Then there must be some firm $i$ s.t. $n'_i < n_i$ (i.e. with fewer workers). In particular, $m_i(1,n'_i) \geq m_i(1,n_i-1)$. By Lemma~\ref{lemma:marginal}, $m_i(1,n_i-1)\geq \delta^*$.

\subpar{Case~2b} $P'= P^*$.
By the minimality of $\delta^*$, there is a firm $i$ s.t. $m_i(1,n_i)\geq \delta^*$ (and this is in fact an equality by Lemma~\ref{lemma:marginal}). 
 It thus holds for $i$ that $m_i(1,n'_i) = m_i(1,n_i) \geq  \delta^*$.

 Since in either case there is a firm s.t. $m_i(1,n'_i)\geq \delta^*$ (w.l.o.g. firm~1), we apply the following argument.
  Firm~$1$ can re-deviate from $(P',\vec x')$ by offering $x''_{1,j^*} = \delta^*-\eps/2$ to a worker $j^*\in A_2$. Observe that  $x''_{1,j^*} > \delta^*-\eps\geq \hat x'_{j^*} = \max_{i>1} x'_{i,j^*}$. Since $j^* \notin R$, her strategy dictates she will now join firm~1. Thus in the newly formed partition $P''$,
 $$r''_1 = r'_1 +m_1(1,n'_1)-x''_{1,j^*} \geq r'_1 +\delta^* - (\delta^* -\eps/2) > r'_1,$$
 i.e. firm~$i$ has an incentive to depart from the coalition $R$.
\end{proof}

\paragraph{Non-homogeneous games}
If we restrict ourselves to \emph{proportional} policies,  It can be shown (see Example~\ref{ex:no_prop_CRP} in the appendix) that coalition-proofness, or even cartel-proofness, is unattainable. This holds even in the case of two firms, where we know a proportional PSPE must exist. We conjecture that a CP-PSPE may not exist for two firms even if we relax the proportionality requirement.


\subsection{Revenue distribution}
\label{sec:weighted_revenue}
In cases where an equilibrium exists, we are interested in how revenue is distributed between firms and workers, focusing on symmetric weighted games. We derive a closed formula for firms' revenue in games where we know a PSPE exists, and test how well it generalizes to other weighted games.

We define a baseline estimate for a firm's revenue as follows.
\begin{align*}
&r_0 \stackrel{\text{def.}}{=} v(q)-\delta\cdot q,~~\text{where } \\
&q = \floor{w(N)/k}, \delta=\frac{\max_i v(S_i)-\min_i v(S_i)}{\max_i w(S_i)-\min_i w(S_i)}.
\end{align*}

\paragraph{Two symmetric firms}
Suppose that $w(S_1)\geq w(S_2)$. Then in the proportional equilibrium we get that the payoff per unit of weight is $\delta = \frac{v(S_1)-v(S_2)}{w(S_1)-w(S_2)}$ as a special case of Theorem~\ref{th:PSPE_fair_k2_distinct}. 
 Both firms get a revenue of $r=v(S_1)-\delta w(S_1) = v(S_2)-\delta w(S_2)$. Note that as $w(S_1),w(S_2)$ become closer, $r$ becomes closer to $r_0$.

\paragraph{Almost-Balanced partitions}
Indeed, in cases where an almost-balanced partition exists (and in particular in homogeneous games), it can be shown (see proof of Prop.~\ref{th:PSPE_balanced}) that the marginal value of a unit of weight is $\delta = m(1,q)$. Then the revenue of each firm is either $v(q)-q\delta=r_0$, or $v(q+1)-(q+1)\delta = v(q+1)-m(1,q) -q\delta=r_0$. That is, it exactly equals our baseline prediction.
$\delta$ can be thought of as the slope, or ``derivative'' of $v$ around the balanced point $q$. Thus as $v$ is ``more concave'', the marginals decrease faster, the payments to workers are lower, and the profit of the firms increases. As a concrete example, if $v(w)=w^\alpha$ (for $0<\alpha<1$), then $\delta \approx v'(q)=\alpha\cdot q^{\alpha-1}$ and 
$r_0 \approx q^\alpha - q \cdot \alpha\cdot q^{\alpha-1} = (1-\alpha)q^\alpha = (1-\alpha) v(q)$,
i.e. the firms keep a fraction of $1-\alpha$ from the total value, and the workers share the complement fraction of $\alpha$. 


Thus for both cases where the existence of equilibrium is guaranteed (two firms and almost-balanced partitions), we have a closed form for firms' revenue in the ``natural'', proportional, equilibrium. 


\paragraph{Three symmetric firms}
Since with three firms there is no PSPE that can be described in a closed form, and sometimes no PSPE at all, we looked on the entire range delimited by the lowest and highest revenue attainable in any PSPE.
For each of the randomized games generated in Section~\ref{sec:weighted_k}, and using the linear program $LP(G,P)$ described in Section~\ref{sec:LP}, we computed for every instance the PSPEs that yield the minimal and the maximal revenue for the firms, respectively (ignoring the few cases where a PSPE did not exist). We then compared these values to our heuristic revenue prediction according to the baseline $r_0$ described above. 

It turns out that the heuristic prediction is quite accurate. When using the value function $v(w)=w^\alpha$ (for which $r_0=1-\alpha$), the revenue was typically within $r_0\pm 10\%$. When using a random value function there was a wider fluctuation of revenue, but the baseline prediction $r_0$ was almost always within the range of observed values. For more details see Appendix~\ref{sec:emp_revenue}.


\section{Games over a social network}
\label{sec:network}
We next turn to study a second class of games, where the workers are a set of influential nodes in a social network.

\subsection{Network model}
Consider a social network $H=\tup{V,E_H}$ (a directed graph), and a subset of ``influencers'' $N\subseteq V$. Given some diffusion scheme in the network, every set  $S\subseteq N$ influences some portion of the nodes $V$, whose size is denoted by $I_H(S)$.

Given such a social network $H$ and a set of firms $K$, we define a symmetric competition game where firms recruit influencers, trying to advertise to as many people (nodes of $H$) as possible.  The value function of every firm is thus $v_i(S)=v(S)=I_H(S)$.\footnote{In principle, we can also define asymmetric games by using a different network $H_i$ for each firm. 
}

We apply one of the most widely known diffusion schemes, called the \emph{independent cascade model}, which has been suggested by Kempe et al.~\shortcite{KKT03}. We briefly describe the diffusion process; for the full details see the paper by Kempe et al. 




In the Independent Cascade model, every edge in the network $H$ has an attached \emph{probability} $p_{u,u'}$. Once a node $u$ is activated, it tries to activate once each neighbor $u'$, and succeeds w.p. of $p_{u,u'}$, independently of the state of any other node. Once a node is activated, it remains active. 
The influence of a set $S$, denoted by $I_H(S)$, is the expected number of nodes that end up as active if we activate the set $S$. 
{\small
$$v_i(S)=v(S)=I_H(S) = \sum_{u\in V} pr(u \text{ is activated} | S \text{ is active}).$$
}
This is equivalent to summing the probabilities of all percolations (subgraphs of $H$) in which there is a directed path from some node in $S$ to $u$. We should note that $I_H(S)$ is a submodular function~\cite{KKT03}. However, $I_H(S)$ does not necessarily hold the gross-substitute condition (see Appendix~\ref{sec:GS}).

While the independent cascade model seems to be more powerful than the weighted model studied in the previous section, it turns out that no model generalizes the other. Indeed, Example~\ref{ex:weighted_no_network} in the appendix demonstrates a weighted value function over $3$ homogeneous workers, that cannot be represented as the influence in any graph $H$. 
Therefore, weighted value functions and influence value functions are two different classes of submodular valuations. A natural question is whether a PSPE always exists with two firms in the independent cascade model. Unfortunately, the answer is negative in the general case (as we show later in Section~\ref{sec:dense}).
 Nevertheless, we can show a weaker result by adding constraints on the network structure.

\medskip
We say that a network $H=\tup{V,E_H}$ is $t$-sparse (w.r.t the set $N\subseteq V$), if every node $u\in V$ can be reached by at most $t$ workers from $N$. 

Intuitively, $t$-sparsity means that the influence cones of different workers hardly intersect. A $1$-sparse network means that the cones of influence are pairwise mutually exclusive and thus that the influence is completely additive (a trivial case). A $2$-sparse network means that two cones may intersect, but never three or more. An $n$-sparse network puts us back in the general case. In order to analyze sparse networks, it will be useful to formally define synergy graphs.

\subsection{Synergy graphs and sparse social networks}
\label{sec:sparse}

A \emph{synergy graph} is an undirected graph $M=\tup{N,E_M}$ with non-negative weights, where self-edges are allowed. It can thus be represented as a symmetric matrix, which is also denoted by $M$. Every synergy graph $M$ induces a value function $v_M$, where the value of a coalition is the sum of weights of edges between coalition members (including self edges), and edges going outside the coalition. That is, 
$$v_M(S) = \sum_{j\in S}M(j,j) + \sum_{j,j'\in S, j< j'} M(j,j') + \sum_{j\in S} \sum_{j''\notin S}M(j,j'').$$

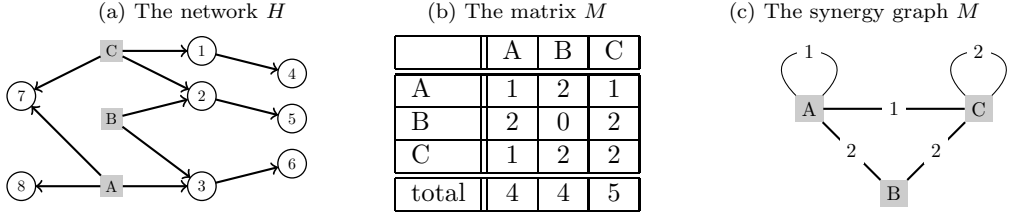
\begin{figure*}
\centering
\subfloat[The network $H$]{\label{sfig:H}
\begin{tikzpicture}[scale=0.6,transform shape]

  \Vertex[x=2,y=4]{1}
  \Vertex[x=2,y=3]{2}
  \Vertex[x=2,y=1]{3}
  \Vertex[x=4,y=3.5]{4}
  \Vertex[x=4,y=2.5]{5}
  \Vertex[x=4,y=1.5]{6}
  \Vertex[x=-2,y=3]{7}
  \Vertex[x=-2,y=1]{8}
  \tikzstyle{VertexStyle}=[fill=black!20!white]
  \Vertex[x=0,y=1]{A}
  \Vertex[x=0,y=2.5]{B}
  \Vertex[x=0,y=4]{C}
  
  \tikzstyle{LabelStyle}=[fill=white,sloped]
  \tikzstyle{EdgeStyle}=[->]
  \Edge[](C)(1)
  \Edge[](C)(2)
  \Edge[](1)(4)
  \Edge[](B)(2)
  \Edge[](B)(3)
  \Edge[](A)(3)
  \Edge[](2)(5)
  \Edge[](3)(6)
  \Edge[](A)(7)
  \Edge[](C)(7)
  \Edge[](A)(8)
\end{tikzpicture}
~~~~~~~
}
\subfloat[The matrix $M$]{\label{sfig:M}
\begin{tabular}{|l||c|c|c|}
\hline
    & A & B & C \\
\hline
\hline
 A  & 1 & 2 & 1 \\
 \hline
 B &  2 & 0 & 2 \\
 \hline
 C & 1 & 2 & 2 \\
 \hline
 \hline
 total & 4 & 4& 5 \\
 \hline
\end{tabular}
}
\subfloat[The synergy graph $M$]{\label{sfig:MG}
~~~~~~~~
\begin{tikzpicture}[scale=0.75,transform shape]
 \tikzstyle{VertexStyle}=[fill=black!20!white]
 \tikzstyle{EdgeStyle}=[-]
  \tikzstyle{every node}=[node distance = 4cm,%
                          bend angle    = 45,%
                          fill          = gray!30]
  \Vertex[x=0,y=0]{A}
  \Vertex[x=3,y=0]{C}
  \Vertex[x=1.5,y=-1.5]{B}
  \Edge[label=$2$](A)(B)
  \Edge[label=$2$](B)(C)
  \Edge[label=$1$](A)(C)

\tikzset{every loop/.style={}} 
  \tikzstyle{every node}=[node distance = 4cm,%
                          bend angle    = 45,%
                          fill          = white!30]
  
  \path  (A)   edge[loop] node  {$1$} (A); 
  \path  (C)   edge[loop] node  {$2$} (C); 
\end{tikzpicture}
}
\caption{\label{fig:network}An example of a sparse network / synergy graph with three workers $A,B$ and $C$. The maximal cut in $M$, which is also the optimal partition, is $P^*=(\{A,C\},\{B\})$. Then $SW(P^*) = v(\{A,C\}) + v(\{B\}) = 8 + 4  =12$. 
}
\end{figure*}

\begin{lemma}
\label{th:synergy_influence}
A value function $v$ over a set of workers $N$ can be described by a synergy graph if and only if it can be described as the influence in a $2$-sparse network. I.e. there is $M$ s.t. $v=v_M$ iff there is $H$ s.t. $v=I_H$.
\end{lemma}
As an intuition, the mapping is constructed s.t. $M(j,j')$ equals the expected number of nodes that are influenced by \emph{both} worker $j$ and worker $j'$ (i.e. the intersection of their influence cones).
Fig.~\ref{fig:network} demonstrates a network $H$ and its corresponding synergy graph $M$.
Our main positive result in the network model is the following.

\begin{theorem}
\label{th:PSPE_sparse}
Let $G=\tup{N,K,v_M}$ be a symmetric competition game with 2 firms over a synergy graph $M$.
Then $G$ has a  PSPE.
\end{theorem}
The outline of the proof is as follows.  The maximal cut in $M$ is the optimal partition in $G$, and we set the payoff of each worker $j$ to be the average of her total influence and her exclusive influence (i.e., $x_j = \frac{v(\{j\}) + M(j,j)}{2}$), and show that $\vec x$ induce a PSPE. 
On the other hand, the above result does not extend to games with more than two firms:

\begin{proposition}
\label{th:no_PSPE_2_sparse_k3}
There is a symmetric competition game with 3 firms over a synergy graph, that has no PSPE.
\end{proposition}
For the proof of the proposition, see Example~\ref{ex:no_PSPE_2_sparse_k3} in the appendix.
From the last two results and Lemma~\ref{th:synergy_influence}, we have that a PSPE always exists in symmetric games over $2$-sparse networks when there are two firms, but not with three or more.

\subsection{Dense and asymmetric networks}
\label{sec:dense}
What about asymmetric firms, or more influence networks that are not 2-sparse? Consider the following example by Dobzinski and Schapira~\shortcite{DobS06} of a submodular combinatorial auction with two bidders and $4$ items.

\begin{example}[\cite{DobS06}]
\label{ex:dense_DS}
The first valuation function $v_1$ is defined as follows:

The value function $v_2$ is defined similarly, replacing $\{1,3\}$ and $\{2,4\}$ with $\{1,2\}$ and $\{3,4\}$. Dobzinski and Schapira show that this game has no equilibrium.
\end{example}
$$v_1(S)=\left\{ \begin{array}{ll} 
2, & |S|=1 \\
3, & |S|=2, S\neq \{1,3\} \text{ and } S\neq \{2,4\} \\
4, &S = \{1,3\} \text{ or } S = \{2,4\} \\
4, & |S|\geq 3
 \end{array}\right. $$


We next use Example~\ref{ex:dense_DS} as a starting point, and further develop it do derive negative results for competition games over asymmetric and dense influence networks.
 

We show how to construct a particular network where the influence equals the production function $v$. We first describe a single network that equals $v_1$. The network $H_1$ contains our four workers $W=\{1,2,3,4\}$ and four other nodes $A=\{a_{1,2},a_{2,3},a_{3,4},a_{4,1}\}$. To every $a_{i,j}$, there are edges with influence probability $1$ both from $i$ and from $j$. It is not hard to see that for every $S\subseteq A$, $I_{H_1}(S)=v_1(S)$. Note that $v_2$ can be constructed in a similar way by a network $H_2$.
\begin{corollary}[From Example~\ref{ex:dense_DS}]
\label{th:no_PSPE_sparse_distinct}
There is an asymmetric competition game with two firms, each over a different 2-sparse network, that has no PSPE.
\end{corollary}

\newpar{Symmetric games} The following lemma describes a ``symmetrization'' that can be applied on any submodular game with two firms. 
\begin{lemma}
\label{lemma:sub_sym}
Let $G=\tup{N,\{1,2\},v_1,v_2}$ be a submodular game with two firms. Let $Z =  \max\{v_1(N),v_2(N)\}, Z'>Z$, and $N'=N\cup\{x,y\}$. Define a new value function $v$ s.t. for every $S\subseteq N$, 

\vspace{-4mm}
$$\begin{array}{ll}
  v(S) = v_1(S)+v_2(S) &  v(S_x) = v_1(S) + Z + Z' \\
 v(S_y) = v_2(S) + Z + Z' ~~~~& v(S_{x,y}) = 2Z + Z',  
\end{array}
$$
where $S_x = S \cup \{x\}$, and likewise for $y$ and $\{x,y\}$.
Then the symmetric game $G'=\tup{N',\{1,2\},v}$ is submodular. Further, $G'$ has a PSPE iff $G$ has a PSPE.
\end{lemma}

Next, we apply such a symmetrization to construct the full network $H$. $H$ has the set of workers $\{1,2,3,4,x,y\}$. We have three sets of nodes: $A$ is defined as above, $B=\{b_{1,3},b_{3,2},b_{2,4},b_{4,1}\}$, and another set $C$ of size $Z'>Z=4$.
We add edges from $W$ to $A$ as in $H_1$ and to $B$ as in $H_2$. In addition, $x$ is connected to all of $B \cup C$, and $y$ is connected to all of $A \cup C$. It is easy to see that for all $S\subseteq \{1,2,3,4\}$, $v(S_x) = v_1(S) + |C| + |B| = v_1(S) + Z' +Z = v(S_x)$, and similarly for $y$. Also, $v(S_{x,y}) = |A|+|B|+|C| = 2Z + Z'$, and $v(S) = v_1(S)+v_2(S)$, as required in Lemma~\ref{lemma:sub_sym}.

Note that every node in $H_1,H_2$ is affected by exactly two workers, and every node in $H$ is affected by three. We get the following result as a corollary. Together with Corollary.~\ref{th:no_PSPE_sparse_distinct} above it shows that the conditions in Theorem~\ref{th:PSPE_sparse} are minimal.
\begin{proposition}
\label{th:no_PSPE_3_sparse}
	There is a competition game over a $3$-sparse network $H$ with two identical firms and no PSPE.
\end{proposition}



\section{Discussion}
\label{sec:discussion}
We considered combinatorial labor markets without gross substitutes, and showed that a pure subgame perfect Nash equilibrium is guaranteed to exist under certain restrictions, with a special focus on the case of few firms.
 
 In games based on capacities (weights) with subadditive production, we proved that a  PSPE must exist if there are two firms or if there is an almost balanced partition, and demonstrated empirically that it often exists even when neither  condition applies. 

When considering stability against deviations of a coalition, we showed that whenever a PSPE exists, then there is also a cartel-proof-PSPE. 
When workers are homogeneous we proved that there is always a coalition-proof-PSPE. This  means that even if some workers choose to collaborate with a cartel of firms, then there is at least one agent---either a firm or a worker---that will be better off outside this group.
         
 As for the profit distribution, we offered a baseline prediction where salaries depends on the marginal contribution of one unit of weight, assuming weights are equally divided. We showed that this prediction is accurate whenever an almost-balanced partition exists, and gives a good estimation in other cases.
					
Finally, we showed that a PSPE always exists in a particular case of the network model, when the network is sparse and featuring two identical competing firms. Unfortunately, there may not be a PSPE if any of these conditions is violated.

Our results are summarized in Table~\ref{tab:exist}.

\paragraph{Related work and implications}
\label{sec:related}


An obvious difference between our setting and the KC model~\cite{KC82} is that we studied value functions that do not hold the gross-substitutes condition. 
More fundamentally, Kelso and Crawford focused on the question of dynamics and convergence. In contrast, we proposed a simpler two-step process, whose equilibria (PSPE) are defined w.r.t. deviations of a single agent, but turns out to coincide with Kelso and Crawford's core. 
In addition, we analyze stability to group deviations, which becomes non-trivial in our model. 

In the original KC model, workers can express preferences over firms. We conjecture that some of our results on the existence of PSPE can be extended  to this more general model, but not the results on coalitional stability. This is since the latter leans on symmetry between workers, which is broken once preferences are introduced.

The inverse economic model, in which firms compete over \emph{consumers}, rather than workers, is known as \emph{Bertrand competition}~\cite{Bert83}. 
Such competitions have been extended to combinatorial domains by Chawla and Roughgarden~\shortcite{CR08}, 
which described a particular  two-step interaction between sellers (the firms) and consumers. 
 While  Chawla and Roughgarden do not explicitly refer to subgame perfection, the solution concept they apply coincides with PSPE.   

Following Kelso and Crawford's proof that every set of firms (or bidders in combinatorial auctions) whose value functions hold the gross substitutes property, Gul and Stacchetti~\shortcite{GulS99} studied the implications of this restriction. They showed that only a tiny fraction of all submodular value functions are gross substitutes. Further, they proved that for \emph{any} value function without this property, it is possible to construct a market with no equilibrium. However, the construction by  Gul and Stacchetti used an unbounded number of firms/bidders (one for every worker/item). Our results demonstrate that there are value functions where the existence of equilibrium can be guaranteed if the number of firms is low. In particular, it follows from our results that a Walrasian equilibrium always exists for two bidders, whose value functions are either weighted or based on influence in a sparse network.

Dobzinski and Schapira~\shortcite{DobS06} study upper and lower bounds on the integrality gap of various submodular value functions. The integrality gap is an important factor in the construction of efficient approximation algorithms that find optimal allocations in combinatorial auctions. For general submodular functions, they show that the (maximal) integrality gap is between $\frac87$ and $\frac43$. Since we essentially use the same construction, we get that the lower bound of $\frac87$ still applies for value functions based on sparse networks. As for weighted functions, the integrality gap of Example~\ref{ex:sym_4_no_PSPE} is $\frac{47}{46}$, which gives us a lower bound.  An interesting challenge is to find the maximal integrality gap of instances that correspond to weighted or network games. In particular, it is an open question whether tighter bounds can be proved w.r.t. general submodular functions.

\begin{table*}[t]

\begin{tabular}{|l||c|c|c|c|c|}
\hline
                & homogeneous & \multicolumn{2}{|c|}{weighted games} & \multicolumn{2}{|c|}{network games} \\
   \# of firms  & games & near-balanced& non-balanced & $2$-sparse (syn. graphs) &  non-$2$-sparse \\
\hline
\hline
 $k=2$, symm.     & \multirow{4}{*}{+} 
 & \multirow{2}{*}{V ($\Leftarrow$)} & \multirow{2}{*}{V (T.~\ref{th:PSPE_fair_k2_distinct})}    & V (T.~\ref{th:PSPE_sparse}) & X (P.~\ref{th:no_PSPE_3_sparse})\\
 \cline{5-6}
 $k=2$, asym.    &  V (KC~~~\shortcite{KC82}) &  &   & X (C.~\ref{th:no_PSPE_sparse_distinct}) & \multirow{3}{*}{X ($\Rightarrow,\Downarrow$)} \\
 \cline{3-5}
 $k\geq 3$, symm.     &   \multirow{2}{*}{CP-PSPE (P.~\ref{th:CP_PSPE})}           &   V (P.~\ref{th:PSPE_balanced}a)  & X  (P.~\ref{th:no_PSPE_k3})&  X (P.~\ref{th:no_PSPE_2_sparse_k3}) & \\
 \cline{3-5}
 $k\geq 3$, asym.   &             &  X (P.~\ref{th:PSPE_balanced}b)   & X  ($\Downarrow$)& X ($\Downarrow$) & \\
 \hline
\end{tabular}
\caption{\label{tab:exist}
Existence results for the cases where a PSPE is guaranteed to exist are marked with V. Cases marked with X mean that there is an instance where \emph{no PSPE} exists. 
}
\end{table*}

\paragraph{Possible extensions}
In our model firms compete only on the services of the workers, in separation from other arenas in which they might affect one another. 
However, if companies are also competing for market share, then there are externalities:  users in the network that are exposed to an ad of one company may become less likely to purchase the products of its competitor. Ways to handle  externalities between coalitions have been studied in the cooperative games literature (see e.g.~\cite{Bloch96}).


Other future extensions may consider more flexible payoff schemes, such as those considered by Chatterjee et al.\shortcite{Chat+93}. That is, the payment that a firm promises to a given worker might be conditioned on the generated value, or even on the particular set of workers that select the firm. 
Allowing more flexible strategies may expand the set of equilibria, perhaps guarantying existence of PSPE in cases where it is not attainable using fixed payoffs. 




\bibliographystyle{plain}
\bibliography{competition}

\appendix
\section{Proofs}
\label{sec:proofs}
We provide proofs for propositions in the order they appear in the main text. Proofs for each Section~X appear in Appendix~\ref{sec:proofs}.X.
\addtocounter{subsection}{1}
\subsection{Preliminaries}
\begin{rlemma}{lemma:cartel_proof}
Let $G=\tup{N,K,(v_i)_{i\in K}}$ be a competition game s.t. $G$ has at least one PSPE. Let $(P,\vec x)$ be a PSPE of $G$ which minimizes the sum of workers' payoffs $x(N)$. Then $(P,\vec x)$ is also cartel-proof.
\end{rlemma}
\begin{proof}
Let $K'$ be a coalition of  firms, and suppose that there is a deviation to another outcome $(P', \vec x_{K\setminus K'}, \vec X_{K'})$. 
Suppose first that $K'\neq K$, i.e. at least one firm is not part of the coalition $K'$. 
Consider any $i\in K'$. Since there is at least one firm that keeps the original strategy, offering $x_j$ to each worker $j$. Therefore, $i$ must pay at least $x_j$ to every worker $j\in S'_i$ (regardless of the strategies of $K'\setminus \{i\}$). Therefore if $r'_i <r_i$, then $i$ also has a unilateral deviation from $(P,\vec x)$. However this is impossible as $(P,\vec x)$ is a PSPE.

The second case is where $K=K'$, i.e. all firms collaborate. Denote by $\hat{\vec x}'$ the realized payments then are induced by the new outcome $(P',\vec X')$. Since $SW(P)\geq SW(P')$, and $r'_i > r_i$ for all $i\in K$, it must be the case that $\hat x'(N) = SW(P')-\sum_{i\in K} r'_i < SW(P) - \sum_{i\in K} r_i = x(N)$. Thus by our selection of $(P,\vec x)$, we have that $(P',\vec X')$ is \emph{not} a PSPE. In particular, either there is a worker $j$ who prefers to move (in which case we are done), or there is a firm~$i$ and an alternative policy $\vec x''_i$ leading to an outcome $(P'',(\vec X'_{-i},\vec x''_i))$, s.t. 
$r''_i = v_i(S''_i) - \sum_{j\in S''_i} x''_{i,j} > r'_i$.

Thus the coalition $K$ is unstable, and $(P,\vec x)$ is cartel-proof.
\end{proof}

\subsection{Properties}

\begin{rlemma}{lemma:marginal}
Let $(P,\vec x)$ be a PSPE outcome in game $G$. Then for all $i\in K$ and $j\in S_i$,
\begin{enumerate}
    \item $ x_j \leq m_i(j,S_i\setminus \{j\})$.
    \item For any $i'\neq i$, $ x_j \geq m_{i'}(j,S_{i'})$.
\end{enumerate}
\end{rlemma}
\begin{proof}
For the first part, note that if $i$ pays $j$ more than its current marginal value, then $i$ can gain by dismissing $j$.
For the second part, if $i$ pays $j$ less than its marginal value to $i'$, then $i'$ can gain by offering $j$ slightly more than its current payment. All other payments remain unchanged, thus the new outcome will be identical to $P$, except $j$ will join $i'$ instead of $i$.
\end{proof}

\begin{rlemma}{lemma:fair}
If $(P,\vec x)$ is a PSPE in game $G$, then there is a \emph{fair} outcome $(P,\vec x^*)$ that is a PSPE in $G$, where the profit of each firm remains the same.
\end{rlemma}
\begin{proof}
Let $(P,\vec x)$ be a  PSPE, and suppose that workers $j,j'$ are of the same type, $j\in S_i$ and $j'\in S_{i'}$.
If $i\neq i'$, then $j,j'$ must be paid the same. Otherwise, suppose that $x_j < x_{j'}$. firm $i'$ will dismiss $j'$ and recruit worker $j$ instead by offering him $x_j+\eps$ (which means that $(P,\vec x)$ is not stable). 

If $i=i'$, then w.l.o.g. $i$ can pay $j,j'$ the same by equalizing the payment. That is, there is another  PSPE $(P,\vec x^*)$, where $x^*_j = x^*_{j'} = \frac{x_j +x_{j'}}{2}$.
\end{proof}

\paragraph{Computation}

\begin{proposition}\label{th:compute_types}
Let $G$ be a competition game with a $t$ types of workers, where $t$ is fixed. 
\begin{itemize}
\item Given an optimal partition $P$, with any number of firms,  we can find in polynomial time a payment vector $\vec x$ s.t. $(P,\vec x)$ is a PSPE of $G$ (or return that one does not exist).
\item If the number of firms $k$ is also fixed, then a  PSPE $(P,\vec x)$ can be computed in polynomial time (or return that one does not exist).
\end{itemize}
\end{proposition}
\begin{proof}
 Suppose that there is only a fixed number of types $t$, with $n_j$ workers of each type.  Then there are $\prod_{j\leq t}(n_j+1) \leq \(\frac{n}{t}+1\)^t$ types of subsets. Moreover, in a fair PSPE agents of the same type are paid the same. We can therefore write and solve a much smaller linear program, with $t$ variables and at most $k\cdot (n/t+1)^t$ constraints. Since a linear program can be solved in time that is polynomial in its size, the first part of the proposition is settled.

Having a limited number of types also enables us to efficiently find the optimal partition, if the number of firms is fixed as well. 
To see this, note that there are ${n_j+k \choose k}$ ways to partition all workers of type $j$ among the firms. Since we can independently partition each type, we have that the total number of distinct partitions is
\begin{align*}
&\prod_{j\leq t}{n_j + k \choose k} \leq \prod_{j\leq t}\(\frac{e(n_j+k)}{k}\)^k \\
&= e^k\prod_{j\leq t}\(\frac{n_j}{k}+1\)^k \leq e^k\prod_{j\leq t}\(\frac{n}{tk}+1\)^k = e^k\(\frac{n}{tk}+1\)^{tk}.
\end{align*}
Together with the first part of the proposition, this gives us an algorithm to find $(P,\vec x)$, by first generating an optimal partition, and then solve the linear program corresponding to this partition.
\end{proof}

\paragraph{Efficient computation in weighted games}
\label{sec:weighted_compute}

The algorithm above shows that if both the number of firms $k$ and the number of worker types $t$ are fixed, then we can compute a  PSPE in polynomial time, or return that such a solution does not exist (for any competition game). 

However, the time to find the optimal partition of workers $P$ is exponential in $k\cdot t$, making the computation infeasible even for fairly low values of $k$ and $t$. As weighted games have a particularly simple structure, we hope that this will allow us a more efficient computation.
In the general weighted case it is computationally hard to find a  PSPE even with two firms, as this is equivalent to the {\sc Partition} problem, which is NP-hard. 
However, it is well-known that the {\sc Partition} problem is in fact efficiently solvable if weights are polynomially bounded. Indeed, we show that a similar positive result is available in our setting as well. 

Note first that in such case the total weight of agents selecting a firm is also bounded, and thus the value functions can be represented as vectors of polynomial length.  Therefore the entire game has a compact representation, for any number of worker types.

\begin{proposition}
\label{th:weighted_compute}
Let $G=\tup{N,K,\vec w,(v_i)_{i\in K}}$ be a weighted competition game with a fixed number of firms $k$, and let $W=w(N)=\sum_{j\in N}w_j$ . Then an optimal partition can be computed in time $O\(n\cdot k \cdot W^k\)$.
\end{proposition}
\begin{proof}
We apply a dynamic algorithm that is similar to the {\sc Partition} algorithm. We first enumerate all the possible ways to divide the total weight $W$ among the $k$ firms, s.t. the total weight of firm~$i$ is $t_i$ (ignoring the actual weights in $\vec w$). There are at most $W^k$ such ways, i.e. polynomial in $n$. 

Next, we create a table with $W^k$ rows and $n$ columns. In every cell $((t_i)_{i\in K},j)$ we write a partition of workers $\{1,2,\ldots,j\}$ to $(S_1,\ldots,S_k)$, s.t. $w(S_i)=t_i$ -- if such a partition exists. We fill the table iteratively from column~1, where for every column $j$, and every non-empty cell $((t_i)_{i\in K},j-1)$, we add worker $j$ to each $S_i$ (if there is more than one partition that fits in a cell we write one arbitrarily). Thus we can fill every column in $O(W^k \cdot k)$ operations, and the whole table in $O(W^k\cdot k\cdot n)$. Finally, to find the optimal partition we compute the value of every non-empty cell in the final column, and select a partition of $N$ that maximizes this value.

Note that when the number of types is limited, then only a small fraction of each column will be filled, and the algorithm becomes even faster.
\end{proof}
A-priory, there may be a large number of optimal partitions, and we must try each one of them to find whether there is a  PSPE or not. In practice, in almost every instance of the problem any optimal partition induces a PSPE, and it is therefore sufficient to compute a single arbitrary optimal partition $P$ and solve its related linear program.
%

\subsection{Weighted games}
\label{sec:proofs_weighted}
%
%

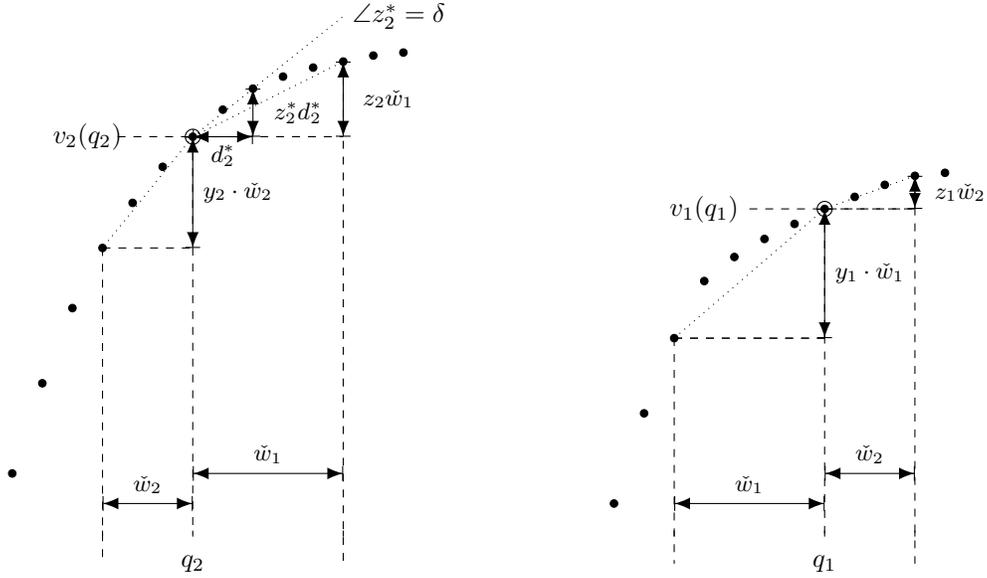
\begin{figure*}[p]
\begin{center}
%

\begin{tikzpicture}[scale=0.4]

\pgfarrowsdeclarecombine{dimarrow}{dimarrow}{latex}{latex}{}{}
\tikzstyle{dot}=[circle,draw=black,fill=black,inner sep=0pt,minimum size=1mm]
\tikzstyle{bigdot}=[circle,draw=black,fill=white,inner sep=0pt,minimum size=2mm]
\def\harrow[#1][#2][#3]{\draw[|-|, 
        decoration={markings, 
                mark=at position .5 with {\node at (0,0.25) {\small{#3}};}, 
                mark=at position 0 with {\arrowreversed[scale=1.5]{dimarrow}};, 
                mark=at position 1 with {\arrow[scale=1.5]{dimarrow}};, 
        }, postaction=decorate] #1 -- #2 };

\def\varrow[#1][#2][#3]{\draw[|-|, 
        decoration={markings, 
                mark=at position .5 with {\node at (0,0.6) {\small{#3}};}, 
                mark=at position 0 with {\arrowreversed[scale=1.5]{dimarrow}};, 
                mark=at position 1 with {\arrow[scale=1.5]{dimarrow}};, 
        }, postaction=decorate] #1 -- #2 };

%
%


%

\node at (0,2) [dot] {};
\node at (1,5)  [dot] {};
\node at (2,7.5) [dot] {};
\path (3,9.5) coordinate (y2);
\node at (y2) [dot] {};
\node at (4,11) [dot] {};
\node at (5,12.2) [dot] {};
\path (6,13.2) coordinate (q2);
\node at (q2) [bigdot] {};
\node at (q2) [dot] {};
\node at (7,14.1) [dot] {};
\path (8,14.8) coordinate (zs2);
\node at (zs2)  [dot] {};
\node at (9,15.2) [dot] {};
\node at (10,15.5) [dot] {};
\path (11,15.7) coordinate (z2);
\node at (z2) [dot] {};
\node at (12,15.9)  [dot] {};
\node at (13,16) [dot] {};

\draw (6,0.1) -- (6,-0.1);
\draw[dashed] (q2) -- (6,0);
\node at (6,-1) {$q_2$};

\harrow[(3,1)][(6,1)][$\check{w}_2$];
\draw (3,0.1) -- (3,-0.1);
\draw[dashed] (y2) -- (3,-1);
\draw[dashed] (y2) -- (6,9.5);
\varrow[(q2)][(6,9.5)][$y_2\cdot \check{w}_2$];

\harrow[(6,2)][(11,2)][$\check{w}_1$];
\draw (11,0.1) -- (11,-0.1);
\draw[dashed] (z2) -- (11,-1);
\draw[dashed] (q2)+(-2.5,0) -- (11,13.2);
\node at (2.5,13.2) {$v_2(q_2)$};
\varrow[(z2)][(11,13.2)][$z_2 \check{w}_1$];

\harrow[(8,13.2)][(q2)][$d^*_2$];
\varrow[(zs2)][(8,13.2)][$z^*_2 d^*_2$];

\draw[dotted] (q2) -- (y2);
\draw[dotted] (q2) -- (z2);
\path (q2)+(5,4.0) coordinate (zs2ex);
\draw[dotted] (q2) -- (zs2ex);
\node[right] at (zs2ex) {$\angle z^*_2 = \delta$};


\node at (20,1) [dot] {};
\node at (21,4)  [dot] {};
\path (22,6.5)  coordinate (y1);
\node at (y1) [dot] {};
\node at (23,8.4) [dot] {};
\draw[dashed] (y1) -- (27,6.5);
\node at (24,9.2) [dot] {};
\node at (25,9.8) [dot] {};
\node at (26,10.3) [dot] {};
\path (27,10.8) coordinate (q1);
\node at (q1) [bigdot]  {};
\node at (q1) [dot]  {};
\node at (28,11.2) [dot] {};
\node at (29,11.6)  [dot] {};
\path (30,11.9) coordinate (z1);
\node at (z1) [dot] {};
\draw[dashed] (q1) -- (30,10.8);
\node at (31,12) [dot] {};

\draw (27,0.1) -- (27,-0.1);
\draw[dashed] (q1) -- (27,0);
\node at (27,-1) {$q_1$};

\harrow[(22,1)][(27,1)][$\check{w}_1$];
\draw (22,0.1) -- (22,-0.1);
\draw[dashed] (y1) -- (22,-1);
\draw[dashed] (y1) -- (27,6.5);
\varrow[(q1)][(27,6.5)][$y_1\cdot \check{w}_1$];

\harrow[(27,2)][(30,2)][$\check{w}_2$];
\draw (30,0.1) -- (30,-0.1);
\draw[dashed] (z1) -- (30,-1);
\draw[dashed] (q1)+(-2.5,0) -- (30,10.8);
\node at (23,10.8) {$v_1(q_1)$};
\varrow[(z1)][(30,10.8)][$z_1 \check{w}_2$];

\draw[dotted] (q1) -- (y1);
\draw[dotted] (q1) -- (z1);


%
%
%
%
%

\end{tikzpicture}
\end{center}
\caption{\label{fig:k2_proof}The value functions of both firms in the partition $P^*$.  Here $\delta = \max\{z^*_1,z^*_2\} = z^*_2$.}
\end{figure*}

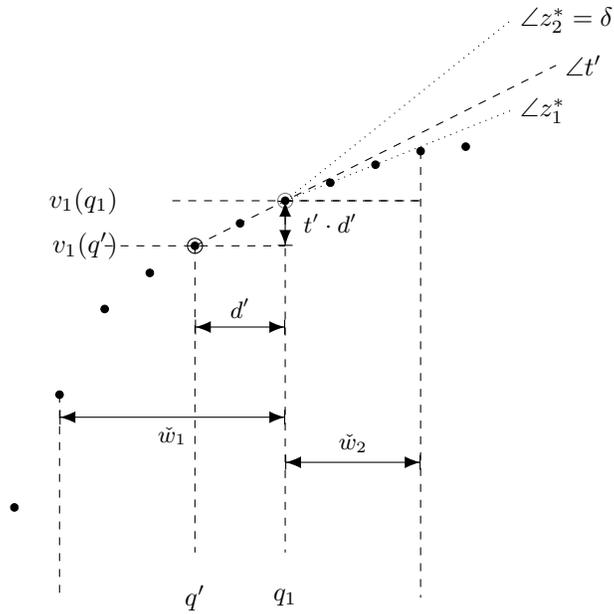
\begin{figure*}[p]
\begin{center}
%

\begin{tikzpicture}[scale=0.6]

\pgfarrowsdeclarecombine{dimarrow}{dimarrow}{latex}{latex}{}{}
\tikzstyle{dot}=[circle,draw=black,fill=black,inner sep=0pt,minimum size=1mm]
\tikzstyle{bigdot}=[circle,draw=black,fill=white,inner sep=0pt,minimum size=2mm]
\def\harrow[#1][#2][#3]{\draw[|-|, 
        decoration={markings, 
                mark=at position .5 with {\node at (0,0.25) {\small{#3}};}, 
                mark=at position 0 with {\arrowreversed[scale=1.5]{dimarrow}};, 
                mark=at position 1 with {\arrow[scale=1.5]{dimarrow}};, 
        }, postaction=decorate] #1 -- #2 };

\def\varrow[#1][#2][#3]{\draw[|-|, 
        decoration={markings, 
                mark=at position .5 with {\node at (0,0.6) {\small{#3}};}, 
                mark=at position 0 with {\arrowreversed[scale=1.5]{dimarrow}};, 
                mark=at position 1 with {\arrow[scale=1.5]{dimarrow}};, 
        }, postaction=decorate] #1 -- #2 };

\node at (1,4)  [dot] {};
\path (2,6.5)  coordinate (y1);
\node at (y1) [dot] {};
\node at (3,8.4) [dot] {};
\node at (4,9.2) [dot] {};
\path (5,9.8) coordinate (q'1);
\node at (q'1) [bigdot] {};
\node at (q'1) [dot] {};
\node at (6,10.3) [dot] {};
\path (7,10.8) coordinate (q1);
\node[circle,draw=gray,fill=white,inner sep=0pt,minimum size=2mm] at (q1)  {};
\node at (q1) [dot]  {};
\node at (8,11.2) [dot] {};
\node at (9,11.6)  [dot] {};
\path (10,11.9) coordinate (z1);
\node at (z1) [dot] {};
\draw[dashed] (q1) -- (10,10.8);
\node at (11,12) [dot] {};

\draw[dashed] (q1) -- (7,3);
\node at (7,2) {$q_1$};

\harrow[(7,6)][(2,6)][$\check{w}_1$];
\draw[dashed] (y1) -- (2,2);

\harrow[(7,5)][(10,5)][$\check{w}_2$];
\draw[dashed] (z1) -- (10,2);
\draw[dashed] (q1)+(-2.5,0) -- (10,10.8);
\node[left] at (3.5,10.8) {$v_1(q_1)$};


\draw[dashed] (q'1) -- (5,3);
\node at (5,2) {$q'$};

\harrow[(5,8)][(7,8)][$d'$];

%

%
%
%

\draw[dashed] (q'1)+(-2,0) -- +(2,0);
\node[left] at (3.5,9.8) {$v_1(q')$};

\varrow[(q1)][(7,9.8)][$t' \cdot d'$];
%
%
%
\path (q1)+(6,3) coordinate (q1ex);
\path (z1)+(3,1.1) coordinate (z1ex);
\path (q1)+(5,2.0) coordinate (zs1ex);
\path (q1)+(5,4.0) coordinate (zs2ex);
\draw[dashed] (q'1) -- (q1ex);
\draw[dotted] (q1) -- (zs1ex);
\draw[dotted] (q1) -- (zs2ex);

\node[right] at (q1ex) {$\angle t'$};
\node[right] at (zs1ex) {$\angle z^*_1$};
\node[right] at (zs2ex) {$\angle z^*_2 = \delta$};

\end{tikzpicture}
\end{center}
\caption{\label{fig:k2_proof_d}A deviation of firm~1 from the outcome $(P^*,\vec x)$ to a policy where $q'=w(S'_1) = q_1-2$.}
\end{figure*}

\begin{rtheorem}{th:PSPE_fair_k2_distinct}
Let $G=\tup{N,K,\vec w,v_1,v_2}$ be a weighted competition game with two firms, then $G$ admits a proportional PSPE.
\end{rtheorem}
\begin{proof}
For any partition $P=(S_1,S_2)$ we denote the social welfare by $SW(P)= v_1(S_1)+v_2(S_2)$. Note that the input $w(S_1)$ contains all the required information to compute $SW$, thus we can write
$$SW(w(S_1)) \stackrel{\text{def.}}{=} v_1(w(S_1))+v_2(W-w(S_1)).$$

Let $P^*=(S^*_1,S^*_2)$ be a partition maximizing the social welfare, and $q_i=w(S^*_i)$.  Let $j'_i = \argmin_{j\in S^*_i} w_j$ and $\check{w}_i = w_{j'_i}$. 

We first handle the case where $q_i=0$ for some $i$, that is where all workers join the same firm. In such case let $w^*=\min_{j\in N}w_j$. By setting the unit payment to $\delta=v_i(w^*)/w^*$, and $\vec x = \delta \cdot \vec w$, we get a PSPE: the marginal value of every set $S$ with weight $w=w(S)\geq w^*$ is at most $x(S)= \delta\cdot w $ to firm $i$, and at least $x(S)$ to firm $-i$. 

Assume therefore that both $S^*_1,S^*_2$ are non-empty. Since value functions are different, the marginal contribution of each unit of weight to each firm may also be different. We therefore replace the ``slope'' $\delta$ with four different quantities $y_1,z_1,y_2,z_2$. 
We denote by $y_i,z_i$ the \emph{normalized} marginal value of the ``lightest'' worker below and above the threshold $q_i$, respectively. Formally, $y_i = \frac{1}{\check{w}_i}(v_i(q_i)-v_i(q_i-\check{w}_i))$, and $z_i=\frac{1}{\check{w}_{-i}}(v_i(q_i+\check{w}_{-i})-v_i(q_i))$.
Now, by subadditivity, $y_i \geq z_i$ for both firms. Also, $y_i\geq z_{-i}$, as otherwise there is a more efficient partition where $-i$ has worker $j'_i$.

Next, set
$$d^*_i = \min\left\{d \geq 1: \frac{v_i(q_i+d) - v_i(q_i)}{d} \leq y_{-i}\right\},$$
and 
$$ z^*_i = \frac{v_i(q_i+d^*_i) - v_i(q_i)}{d^*_i}.$$

Clearly, $y_i,y_{-i}\geq z^*_i\geq z_i$ and $d^*_i \leq w_{-i}$. For a graphical illustration, see Figure~\ref{fig:k2_proof}.

We define a unit payment $\delta = \max\{z^*_1,z^*_2\}$. In particular,  $y_i \geq \delta \geq z^*_i \geq z_i$ for both firms.
We set the (proportional) payment policy so that $x_j = \delta w_j$.


The profit of firm~$i$ is
$$r_i = v_i(S^*_i)-\sum_{j\in S_i}x_j = v_i(q_i) - \delta w(q_i).$$

Assume, toward a contradiction, that some firm can deviate (w.l.o.g. firm~1), and keep a total weight of $q'=w(S'_1)$.

\subpar{case~1} $q'<q_1$. Denote $d'=q_1-q'$, and the marginal value (per unit) of the range $[q',q_1]$ by $t'=\frac{v_1(q_1)-v_1(q')}{d'}$ (see Figure~\ref{fig:k2_proof_d}).

Since firm~1 increased the profit, $t'$ must be strictly smaller than $\delta$, as otherwise $r'_1 = r_1 -d'\cdot t' + d'\cdot \delta < r_1$.
 Thus it is not possible that $d'\geq \check{w}_1$, as this would entail $t'\geq y_1 \geq \delta$.

 On the other hand, $t'\geq z^*_1$ by subadditivity of $v_1$. As $\delta= \max\{z^*_1,z^*_2\}$, and $\delta > t' \geq z^*_1$, we get that $t'< \delta=z^*_2$. Therefore,
{\small
\begin{align*}
S&W(q')  = v_1(q') + v_2(W-q') \\
&= (v_1(q_1)-t'\cdot d') + v_2(q_2+d') \\
& =  (v_1(q_1)-t'\cdot d') + v_2(q_2) + d'\frac{v_2(q_2+d')-v_2(q_2)}{d'} \\
& \geq  (v_1(q_1)-t'\cdot d') + v_2(q_2) + d'\frac{v_2(q_2+\check{w}_1)-v_2(q_2)}{\check{w}_1}  \tag{since $d'<\check{w}_1$ and $v_2$ is subadditive}\\
& = (v_1(q_1)-t'\cdot d') + v_2(q_2) + z^*_2 \cdot d' \\
& = (v_1(q_1)-t'\cdot d') + v_2(q_2) + \delta \cdot d' \\
& > v_1(q_1) + v_2(q_2) = SW(q_1),
\end{align*}
}
in contradiction to the optimality of $P$.

\subpar{case~2} $q'>q_1$. Similarly, we denote $d'=q'-q_1$, and it holds that $d'$ is smaller than $\check{w}_2$. Also, $t=\frac{v_1(q')-v_1(q_1)}{d'}$ must hold $\delta< t'<y_1$.

Now, if $d'\geq d^*_1$, then we have $t' \leq z^*_1$. However this is impossible as $z^*_1\leq \delta < t'$. Thus $d'<d^*_1$, and by definition, $t'>y_2$. Then, similarly to case~1,
\begin{align*}
S&W(q')  = (v_1(q_1)+t'\cdot d') + v_2(q_2-d')  \\
&=  SW(q_1)+t'\cdot d' - d'\frac{v_2(q_2)-v_2(q_2-d')}{d'} \\
& \geq  SW(q_1)+t'\cdot d'  - d'\frac{v_2(q_2)-v_2(q_2-\check{w}_2)}{\check{w}_2}  \tag{since $d'<\check{w}_2$ and $v_2$ is subadditive}\\
& = SW(q_1) + d'\cdot t' - d'\cdot y_2 \\
&= SW(q_1) + d'(t'-y_2) > SW(q_1),
\end{align*}
which again contradicts the optimality of $P$.
\end{proof}

\begin{rprop}{th:PSPE_balanced}
Let $G=\tup{N,K,\vec w,v}$ be a weighted competition game with $k$ symmetric firms. (a) If there is an almost-balanced partition of $\vec w$, then $G$ admits a proportional PSPE. (b) symmetry is a necessary condition.
\end{rprop}

\begin{proof}[Proof of Proposition~\ref{th:PSPE_balanced}a]
Let $W=\sum_{j\in N}w_j$, and $q=\floor{W/k}$. Set $\delta$ to be the marginal gain of a unit weight in an almost-balanced partition, that is, $\delta = v(q +1) - v(q)$.
In the PSPE policy, each firm offers each worker $j$ a payment of $x_j=\delta\cdot w_j$. 

Let $P^*$ be an almost balanced partition, i.e. $w(S_i)=q$ or $q+1$ for all $i\in K$. We argue that $(P^*,\vec x)$ is a PSPE in the suggested policy. Clearly workers have no reason to deviate.

The profit of each firm is
$$r_i = v(S_i) - \delta \cdot w(S_i) = v(q) - \delta q = v(q+1)-\delta (q+1).$$

Suppose some firm $i$ deviates, and ends up with $S'_i$ of weight $q'$. This means that the workers in $S'_i$ are being payed at least $\delta\cdot w(S'_i) = \delta q'$ in total (as otherwise there is a worker $j$ payed less than $x_j$). Thus the profit of $i$ becomes
$$r'_i \leq v(q')-\delta q' \leq r_i,$$
where the last inequality is since $\{q,q+1\}\subseteq\argmax_{q'\in \mathbb N}v(q')-\delta q'$.
\end{proof}
Part (b) of the proposition follows from Example~\ref{ex:asym_balanced}.

\subsection{Networks games}
\label{sec:proofs_network}
%

\begin{rlemma}{th:synergy_influence}
A value function $v$ over a set of workers $N$ can be described by a synergy graph if and only if it can be described as the influence in a $2$-sparse network. Formally, for any $2$-sparse influence network $H$ there is a synergy graph $M$ s.t. $v_M=I_H$; and for any synergy graph $M$ (with integer weights) there is a deterministic $2$-sparse influence network $H$ s.t. $I_H=v_M$.
\end{rlemma}

Note that since we can always multiply $v$ by a positive constant, we can represent any synergy graph with \emph{rational} weights as an influence network.
\begin{proof}
For simplicity, we will consider deterministic networks (where an edge means influence w.p. $1$), and synergy graphs with integer weights. 

``$\Rightarrow$'' Given a network $H=\tup{V,E_H}$ and $N\subseteq V$, we define the following synergy matrix/graph of size $n\times n$. For every $j\in N, u\in V$, we say that $j$ influences $u$ if there is a directed path from $j$ to $u$ in $H$ (recall that influence is deterministic). 

We set $M(j,j')$ as the  number of nodes influenced by \emph{both} $j$ and $j'$. In particular, the weight of the self-edge $M(j,j)$ is the number of nodes that only $j$ influences. Since $H$ is $2$-sparse, every node is influenced by exactly $1$ or $2$ workers, and thus $I_H(j)=M(j,j) + \sum_{j'\neq j} M(j,j') = \sum_{j'\in N} M(j,j')$.  I.e., the sum of its row (or column) in the matrix $M$. For a set $S$, we need to sum the rows of $j\in S$, and then remove all nodes $u\in V$ that have been counted twice, i.e. cells $M(j,j')$ s.t. $j,j'\in S$ and $j\neq j'$. Formally, 
{\scriptsize
$$I_H(S) = \sum_{j\in S} \(M(j,j)  +\frac12 \!\! \sum_{j'\in S\setminus\{j\}}\!\!\!\! M(j,j') +\!\! \sum_{j''\in N\setminus\{S\}} \!\!\!\! M(j,j'')\) = v_M(S).$$
}
Next, we extend the proof to the case of general probabilities on edges. For $j\neq j'$, suppose that a certain node $u\in V$ is influenced by $j$ w.p. $p_j$, and by $j'$ w.p. $p_{j'}$. Then we add $p_j p_{j'}$ to $M(j,j')$ for each such node. Finally, we set $M(j,j)= I_H(j) - \sum_{j'\neq j} M(j,j')$. Note that $M(j,j)$ is exactly the marginal contribution of $j$, i.e. the fraction of nodes that only $j$ influences. Thus $I_H(S)=v_M(S)$ as in the equation above. 

``$\Leftarrow$'' Given a synergy graph/matrix $M$, we define a set of nodes $V_{j,j'}$ for every pair $j,j'\in N$, whose size is $M(j,j')$. Let $V = N \cup \bigcup_{j,j'} V_{j,j'}$. We connect an edge in $H$ from every $j\in N$, to every $u\in \bigcup_{j'\in N}V_{j,j'}$. Thus we have once again $I_H(j) = \sum_{j'\in N} M(j,j') = v_M(j)$, and similarly for sets as in the equation above.
\end{proof}
By applying both directions of the lemma, we get as a corollary that any $2$-sparse network with rational probabilities can be replaced by a deterministic one.

\begin{rtheorem}{th:PSPE_sparse}
Let $G=\tup{N,K,v_M}$ be a symmetric competition game with 2 firms over a synergy graph $M$.
Then $G$ has a  PSPE.
\end{rtheorem}
\begin{proof} 
Every partition of $N$ to the two firms, is a \emph{cut} in $M$. Let $S_1,S_2$ be such a cut. By definition (after rearranging terms), $v(S)=v_M(S)$ equals to  
{\scriptsize
$$\sum_{j\in S} \(M(j,j)  +\frac12 \!\!\sum_{j'\in S\setminus\{j\}}\!\!\!\! M(j,j') +\!\!\sum_{j''\in N\setminus\{S\}}\!\!\!\! M(j,j'')\).$$
}
The last term is the weight of the cut $(S,N\setminus\{S\})$.

 It is therefore easy to see that social welfare is maximized by taking the maximal cut in $M$. We still need to set the payments properly to induce stability.

Let $P^*=(S^*_1,S^*_2)$ be some maximal cut in $M$, and
let $j\in S^*_i$. We set $x_j = \frac12 (v(j)+M(j,j)) = M(j,j) + \frac12\sum_{j'\in N\setminus\{j\}}M(j,j')$. We claim that the payment policy $\vec x$, with $S^*_1,S^*_2$ is a PSPE. 

Indeed, the revenue of either firm is 
{\small
\begin{align*}
r_i& =v(S_i)- \sum_{j\in S_i}x_j =v(S_i)- \sum_{j\in S_i} \(M(j,j) + \frac12\sum_{j'\in N\setminus\{j\}}\!\!\! M(j,j')\) \\
& = \sum_{j\in S_i} \( M(j,j)  +\frac12 \sum_{j'\in S_i \setminus\{j\}}\!\!\! M(j,j') + \sum_{j''\in S_{-i}} \!\!\! M(j,j'')\) - \\
&~~~~~~~~~~ \sum_{j\in S_i} \(M(j,j) + \frac12\sum_{j'\in S_i\setminus\{j\}}\!\!\! M(j,j') +  \frac12\sum_{j''\in S_{-i}}\!\!\! M(j,j'') \)\\
&  = \frac12\sum_{j\in S_i}\sum_{j'\in S_{-i}}M(j,j') = \frac12\sum_{j\in S_1}\sum_{j'\in S_2}\!\!\! M(j,j'),
\end{align*}
}
 i.e. half the weight of the cut $P^*$ (in particular $r_1=r_2$). Suppose that some firm (w.l.o.g. firm~1) changes its policy, which leads to some cut $(S'_1,S'_2)$, and some profit $r'_1$.
 
For the new outcome to be a PSPE, firm~1 must pay at least $x_j$ to  each $j\in S'_1$.

The new profit of firm~$1$ turns out to be at most half the weight of the new cut $P'$:
{\small
\begin{align*}
r'_1 & = v(S'_1)- \sum_{j\in S'_1}x'_j \leq v(S'_1) - \sum_{j\in S'_1}x_j \\
& = \sum_{j\in S'_1} \( M(j,j)  +\frac12 \sum_{j'\in S'_1 \setminus\{j\}}\!\!\! M(j,j') + \sum_{j''\in S_2} \!\!\! M(j,j'')\) - \\
&~~~~~~~~~~ \sum_{j\in S_1} \(M(j,j) + \frac12\sum_{j'\in S_1\setminus\{j\}}\!\!\! M(j,j') +  \frac12\sum_{j''\in S_2}\!\!\! M(j,j'') \)\\
 & = \frac12\sum_{j\in S'_1}\sum_{j'\in S'_2}M(j,j').
\end{align*}
}
However, since $P^*$ is the optimal (heaviest) cut, we get that $r'_1 \leq w(S'_1,S'_2) \leq w(S^*_1,S^*_2) = r_1$, and thus there is no  profitable deviation.
\end{proof}

\begin{rlemma}{lemma:sub_sym}
Let $G=\tup{N,\{1,2\},v_1,v_2}$ be a submodular game with two firms. Let $Z =  \max\{v_1(N),v_2(N)\}, Z'>Z$, and $N'=N\cup\{x,y\}$.  Define a new value function $v$ s.t. for every $S\subseteq N$, 
\begin{align*}
&v(S) &= v_1(S)+v_2(S) \\
&v(S_x) &= v_1(S) + Z + Z' \\
&v(S_y) &= v_2(S) + Z + Z' \\
&v(S_{x,y}) &= 2Z + Z',  \\
\end{align*}
where $S_x = S \cup \{x\}$, and likewise for $y$ and $\{x,y\}$.
Then the symmetric game $G'=\tup{N',\{1,2\},v}$ is submodular. Further, $G'$ has a PSPE iff $G$ has a PSPE.
\end{rlemma}
\begin{proof}
Set $Z''=Z+Z'$.
We first show that $v$ is submodular. Let $S,T \subseteq N$. Clearly $v(S\cup T)\leq v(S)+v(T)-v(S\cap T)$. 
\begin{align}
v(&S_x \cup T) = v_1(S\cup T) + Z'' \notag \\
&\leq v_1(S)+v_1(T)-v_1(S\cap T) +Z'' \label{eq:Sx}\\
	&= v(S_x)+ (v_1(T)-v_1(S\cap T)) \notag \\
	& = v(S_x)+ (v(T_x)-v(S_x\cap T_x)). \notag \\
	v(&S_{x,y} \cup T)  = 2Z+Z' = v(S_{x,y}) \notag \\
	&\leq v(S_{x,y}) + v(T) - v(S_{x,y}\cap T)  \label{eq:Sxy} \\
	v(&S_x) + v(T_y)  = v_1(S) + v_2(T) + 2Z'' \notag \\
	&\geq v_1(S \cap T) + v_2(S\cap T) + 2Z''  = v(S\cap T) + 2Z'',\notag
	\end{align}
	Thus,      
	{\scriptsize
	$$v(S_x \cup T_y) = 2Z+Z' < 2Z'' \leq  v(S_x) + v(T_y) - v(S\cap T).$$
	}
All other cases follow directly from these cases.

Since any PSPE is maximizing the social welfare, $x$ and $y$ must go to distinct firms. This is since for every $S,T\subseteq N$,
$$v(S_{x,y}) + v(T) = 2Z + Z' + v(T) = Z+ v(T_y)  < v(S_x) + v(T_y).$$
Finally, since $v(S_x), v(T_y)$ are just $v_1(S),v_2(T)$ shifted by a constant $Z''$, there is no PSPE in $G=\tup{N,K,v}$. 
\end{proof}

\section{Examples}
\label{sec:examples}
\subsection{The Gross-substitutes condition}
\label{sec:GS}
Informally, the gross-substitutes (GS) condition states that workers can ``substitute'' one another---if the salary of some workers rises and the salary of other remains the same, then a firm would never want to dismiss or replace workers whose salary remains the same~\cite{KC82}. 

\begin{definition}[Gross substitutes~\cite{KC82}] 
A value function $v:2^N\rightarrow \mathbb R_+$ holds the \emph{gross substitutes} condition if the following holds.

Suppose that under prices $\vec x$ the set $S\subseteq N$ maximizes the profit $r(S,\vec x) = v(S)-x(S)$, and let $T\subseteq S$, $\vec x'\geq \vec x$ where $x'_j=x_j$ for all $j\in T$. Then there is a set $S'\subseteq N$ maximizing $r(S',\vec x')$ s.t. $T\subseteq S'$. 
\end{definition}

The following example shows that a weight-based valuation function may not hold the GS condition.
\begin{example}\label{ex:weighted_no_GS}
Consider the valuation function $v(w) = \min\{w,6\}$. We have five workers, two of weight~3 and three of weight~2, i.e. $\vec w=(3,3,2,2,2)$. With the payoff vector $\vec x=\vec w/2$, there are two optimal subsets: $S_1 = \{3,3\}$ and $S_2=\{2,2,2\}$, each yielding a profit of $v(6)-6/2 = 3$. Suppose the firm employs $S_1$, and we now raise the salary of one of the $3$-workers from $x_1=1.5$ to $x'_1=2$. The unique optimal selection under $\vec x'=(x'_1,x_{-1})$ is now $S_2$. This already violates the GS condition, since $x_2$ did not change, yet worker~2 is not part of any optimal solution.

Similarly, if the firm employs $S_2$ and we raise the salary of one of the $2$-workers, then the unique optimal solution $S_1$ does not contain the $2$-workers whose salary remains the same.
\end{example}

We can similarly show that valuation functions defined by synergy graphs/sparse influence networks may not hold the GS condition. Indeed, consider the valuation function $v_1$ from Example~\ref{ex:dense_DS}. Under the payoff vector $\vec x=(1,1,1,1)$ the set $S_1=\{1,3\}$ is optimal. However if the salary of worker $1$ increases, then the unique optimal set is $S_2=\{2,4\}$. $S_2$ does not include worker~$3$ even though $x'_3=x_3$, and thus violates the GS condition.

\subsection{Weighted games}
\label{sec:examples_weighted}
\newpar{A non-subadditive game}
\begin{example}\label{ex:nonsub}
consider a homogeneous and symmetric game with a weighted (non-subadditive) value function $v= (0,3,4,6)$,  two firms, and three workers. 
 Assume towards a contradiction that there is a  PSPE $((S_1,S_2),\vec x)$. W.l.o.g. $\vec x$ is fair (by Lemma~\ref{lemma:fair}). By the FWT, $(S_1,S_2)$ must be an optimal partition, thus one firm (w.l.o.g. firm~1) has two workers, $v(S_1) = v(2)=4, v(S_2)=v(1)=3$. The marginal value of each worker to firm~1 is $1$, thus they are paid \emph{at most} $1$ each. The marginal value of an additional worker to this firm is $2$. Thus firm~2 must pay its worker \emph{at least} $2$. However, by fairness all workers must be paid the same amount. A contradiction. 
\end{example}
\medskip
An experimental analysis shows that in non-subadditive weighted competition games generated at random, no PSPE usually exists (see Appendix~\ref{sec:emp_heuristic}).

\newpar{A symmetric game with three firms}
\begin{example}\label{ex:sym_3_no_PSPE}
We consider the following competition game with 3 firms, which has no  PSPE. The game has 11 workers, with weights $\vec w = (8,8,8,3,3,3,3,3,2,2,2)$. The value function is $v(w) = \sqrt w$. It is easy to verify that there are exactly four optimal partitions  (up to permutations of agents of the same type), in all of which the total weights are $w(S_1)=14$, $w(S_2)=15$, and $w(S_3)=16$ . The partitions are as follows. 
\begin{align*}
&P_1=(\{8,2,2,2\}, \{3,3,3,3,3\},\{8,8\})\\
&P_2 = (\{3,3,3,3,2\},\{8,3,2,2\},\{8,8\})\\
&P_3 = (\{8,3,3\},\{3,3,3,2,2,2\},\{8,8\})\\
&P_4 = (\{8,3,3\},\{8,3,2,2\},\{8,3,3,2\})
\end{align*}

For each of these partitions, Matlab's \verb+linprog+ function returns a set of conflicting constraints in the corresponding linear program. Therefore, none of these optimal partitions can be stabilized, and by FWT there is no  PSPE. 

For higher values of $k$, we can use the same example with additional firms. For each extra firm $i>3$, we add a worker with weight $100$. Thus each additional firm will hire one heavy worker without affecting the competition between the original firms.
\end{example}

\newpar{An asymmetric, balanced game}
\begin{example}\label{ex:asym_balanced}
We define a game with three firms and three workers of weight $11$, and six workers of weight $2$. The value functions are defined as $v_1(w)=\min\{w,20\}; v_2(w) = \min\{w,19\}; v_3(w) =\min\{w,6\}$.

First note that a balanced partition exists, where each firm has one 11-worker and two 2-workers. The optimal partition is $P=(\{11,11\}; \{11,2,2,2,2\}; \{2,2\})$, and has a welfare of 
$$SW(P) = v_1(22)+v_2(19)+v_3(4) = 20+19+4 = 33.$$
(if all three firms have an 11-worker, then $SW\leq 22+12 +6 = 40$)
Assume towards a contradiction that there is a PSPE $(P,(x_2,x_{11})$.

By Lemma~\ref{lemma:marginal}, $x_2=2$. Now, suppose firm~2 recruits an 11-worker instead of all four 2-workers. Then 
$$r_2 = 19-x_{11}-4x_2 \geq r'_2 =19-2x_{11} \Rightarrow x_{11}\geq 4x_2 = 8.$$
On the other hand, if firm~1 recruits three 2-workers instead of an 11-worker, $$r_1 = 20-2x_{11} \geq  r'_1 = 19-x_{11}-3x_3 \Rightarrow x_{11}\leq 1+3x_2 = 7.$$
Thus we have a contradiction.
\end{example}

\paragraph{A game with no proportional cartel-proof PSPE}
\begin{example}\label{ex:no_prop_CRP}
We consider the following competition game with two firms and $4$ workers, with weights ${1,2,3,5}$. There are two optimal partitions, in both of which the there is a firm with total weight 6, that hires the worker~1. By Lemma~\ref{lemma:marginal}, $x_1=1$ in any PSPE $(P,\vec x)$. Thus in any proportional PSPE, $x_j=w_j$ for all $j\in N$. Now, both firms can collaborate by lowering the payoff of workers $3$ and $5$ by $1$, so that $\vec x'=(1,2,2,4)$. The new outcome $(P,\vec x')$ is also a PSPE, thus no firm wants to deviate from it. However, $(P,\vec x')$ is a (group) deviation from $(P,\vec x)$, thus $(P,\vec x)$ is not a coalition-proof, or even cartel-proof.
\end{example}

\subsection{Network games}
\label{sec:examples_network}

There is a weighted value function over $3$ homogeneous workers, that cannot be represented as the influence in any graph $H$. 
\begin{example}\label{ex:weighted_no_network}
Denote by $f(S)$ the expected fraction of nodes in $V$ that is influenced by \emph{all} workers in $S$. For a deterministic network this simply  means the number of nodes $u\in V$ s.t. for every $j\in S$ there is a path from $j$ to $u$. More generally, $f(S)=\sum_{u\in V} f(u,S)$ where $f(u,S)$ is the the total probability of all percolations in which there are paths to $u$ from every $j\in S$ and from no $j'\in N\setminus S$.
It is straightforward to see that $I_{H}(S) = \sum_{T\cap S\neq \emptyset} f(T)$ for all $S\subseteq N$.

Now, consider the weighted value function $v = (0,3,6,8)$. Since $m_v(1,2) = 8-6 = 2$, we have that $f(j) = m(j,\{j',j''\})= m_v(1,2)=2$ for all $j\in\{1,2,3\}$. Next, 
{\small
\begin{align*}
m&(j',j) = I(\{j,j'\}) - I(\{j\}) \\
&= f(j)+f(j')+f(j,j') + f(j,j'') + f(j',j'') + f(j,j',j'') \\
&~~~~~~~~- (f(j)+f(j,j') + f(j,j'')  + f(j,j',j''))\\
& = f(j') + f(j',j'') = 2 + f(j',j''),
\end{align*}
}
thus
$$ 2 + f(j',j'') = m(j',j) = m_v(1,1)  = 6-3 = 3,$$
which entails that $f(j',j'')=1$ for every pair of workers. Finally, 
\begin{align*}
3 &= v(1) = I(\{1\}) = f(1) + f(1,2) + f(1,3) + f(1,2,3) \\
&=  2 + 1 + 1 + f(1,2,3) = f(1,2,3)+4,
\end{align*}
i.e. $f(1,2,3)=-1$. This is a contradiction since every $f(T)$ is a sum of probabilities and thus cannot be negative.
\end{example}

\begin{example}\label{ex:no_PSPE_2_sparse_k3}
Consider a game with three firms, and a set of workers $N=\{a_1,a_2,b_1,b_2,b_3\}$, where in the graph $M$ every pair of workers in connected (with weight $1$), except the pair $(a_1,a_2)$. In the optimal partition (which maximizes the weight of the multicut $P=(S_1,S_2,S_3)$), $a_1$ and $a_2$ must share a firm. Thus w.l.o.g. $S_1=\{a_1,a_2\},\ S_2=\{b_1,b_2\}, S_3=\{b_3\}$, and $SW(P)=6+7+4=17$.

Assume, w.l.o.g., that a PSPE $(P,\vec x)$ exists. Then by fairness all of the $a_i$ workers are payed the same amount $x_a$, and similarly for the $b_i$ workers. Thus we only need to find the values $x_a$ and $x_b$. By envy-freeness, firm~2 and 3 make the same revenue, thus $r_2 = v(S_2)- 2x_b = v(S_3)-x_b$, i.e. $x_b = 7-4 = 3$, and $r_2=r_3=1$. Firm~1 must have the same revenue as well, thus $1=r_1 = v(S_1)-2x_a = 6-2x_a$, and $x_a=2.5$. However, firm~1 has a deviation, by refusing worker $a_2$, and recruit a $b$ worker instead. Then
$$r'_1 = v(\{a_1,b_1\})- x_a-(x_b+\eps) = 7 - 5.5 - \eps > 1 =r_1.$$
\end{example}

Intuitively, it seems that a generalization to $k>2$ can be easily constructed by considering a multicut (rather than a cut). However, this intuition is misleading. 

The following example shows where this intuition fails.
Consider a network $H$, containing three influencers and three nodes with weights $1,2,3$. Each node is influenced by exactly two influencers (w.p. $1$).  The resulting $3\times 3$ matrix induces a graph $M$ that is a triangle, whose edge weights are $1,2$ and $3$ (i.e. all three edges have different weights). If payments are set according to the scheme above then the profit of every firm will be half the weight of the cut \emph{between himself and the others}. Then the firm that recruited the lightest influencer (and hence has the lightest part of the cut) is envious in the other firms.
However, this game does have the following  PSPE: $P=(\{a\},\{b\},\{c\})$, $\vec x = (2,3,4)$, where the profit of every firm is $1$.

\section{Experimental results}
\label{sec:empirical}
We implemented a program that solves any given symmetric weighted game $G$, using Matlab 7. We used a variation of the algorithm above to find an arbitrary optimal partition $P^*$, then applied the Matlab \verb+linprog+ function to solve the induced linear program, i.e. to find a payment vector $\vec x$ s.t. $(P^*,\vec x)$ is a PSPE of $G$ .
 An instance with $k=3$ and $n\leq 15$ is typically solved in less than a second, so that it is possible to collect statistics. In the few cases where a stable payment vector $\vec x$ was not found, we labeled this instance as ``no solution'', without trying any other partition. Thus it is possible that the actual number of instances with no PSPE is even smaller.
 
\subsection{Experimental setting}
\label{sec:emp_setting}
We generated three datasets with the following characteristics.
\begin{itemize}
	\item Dataset $D_1$ had 3000 instances. Each instance had between 5 and 14 workers, divided to 2-4 types. The weight of each type was in the range 2-15, and the total weight was limited to under $80$.  1053 instances  (32\%) had a balanced or nearly balanced partition. In 4 instances the selected partition could not be stabilized with payments (this does not necessarily mean that the instance has no  PSPE, as we did not try other partitions). 
	\item Dataset $D_2$ had 3000 instances. Each instance had between 4 and 11 workers, with weights in the range 2-15, and no restriction on the number of types. The total weight limit was $80$. 1548 instances (52\%) had a balanced or nearly balanced partition. In all instances, the selected partition induced a  PSPE.
	\item Dataset $D_3$ includes the first 200 instances of $D_1$. 
\end{itemize}
In datasets $D_1$ and $D_2$ we used a random subadditive value function for each instance. This is by sampling the marginal values uniformly from $[0,1]$, and then sort them in decreasing order. For $D_3$ we used the value function $f(w) = w^\alpha$, for various values of $0<\alpha<1$. 

Since we only got a handful of instances with no pure equilibrium, we did not make a statistical analysis of these samples. To test the importance of subadditivity, we also generated random value functions without enforcing subadditivity. In this case only 910 from the instances in $D_1$ (27\%) had stable payments (for the selected optimal partition), and slightly more (33\%) in $D_2$. 

\subsection{Experimental validation of the heuristic payment policy}
\label{sec:emp_heuristic}
For every instance in each of the datasets $D_1$ and $D_2$, we computed the heuristic payment vector $x^*$ as follows.
 
Let $P^*=(S_1,S_2,S_3)$ be the optimal partition, $i_+ = \argmax_{i\in K} w(S_{i})$, $i_- = \argmin_{i\in K} w(S_{i})$, then $z_0=\frac{v(w(S_{i_+})) - v(w(S_{i_-}))}{w(S_{i_+})-w(S_{i_-})}$.  In the induced proportional payoff vector, $x^*(j) = z_0\cdot  w_j$ for all $j\in N$. For every instance $G$, we measured the maximal amount a firm can gain by deviating from the profile $(P^*,\vec x^*)$. Formally, $h(G,x^*) = \max_{i\in K, S'\subseteq N} (v(S')-x^*(S')) - (v(S_i)-x^*(S_i))$. The profits are normalized to $1$, so that $h(G,x^*) =0.2$ for example, means that there is a firm in $G$ that can increase its profit by $20\%$ by deviating from $x^*$.

\begin{figure*}[p]
\begin{center}
\includegraphics[width=0.8\textwidth]{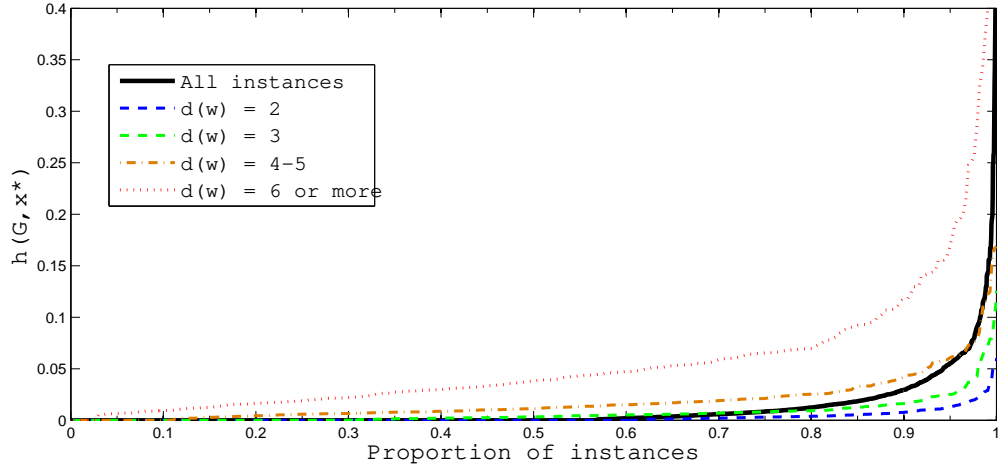}
\caption{\label{fig:D1_x0}  A survival curve of the percentage of instances from the dataset $D_1$, for which $h(G,x^*)\leq h$. The solid line is showing statistics for all instances. The dashed lines are the curves of instances with a particular gap $d=q_{i_+} - q_{i_-}$. We can see that as the gap increases, the stability of the heuristic solution $x^*$ deteriorates. 
We can also see that for almost 60\% of the instances, $h(G,x^*)=0$, i.e. the heuristic solution is stable.
}
\end{center}
\end{figure*}

\begin{figure*}[p]
\begin{center}
\includegraphics[width=0.8\textwidth]{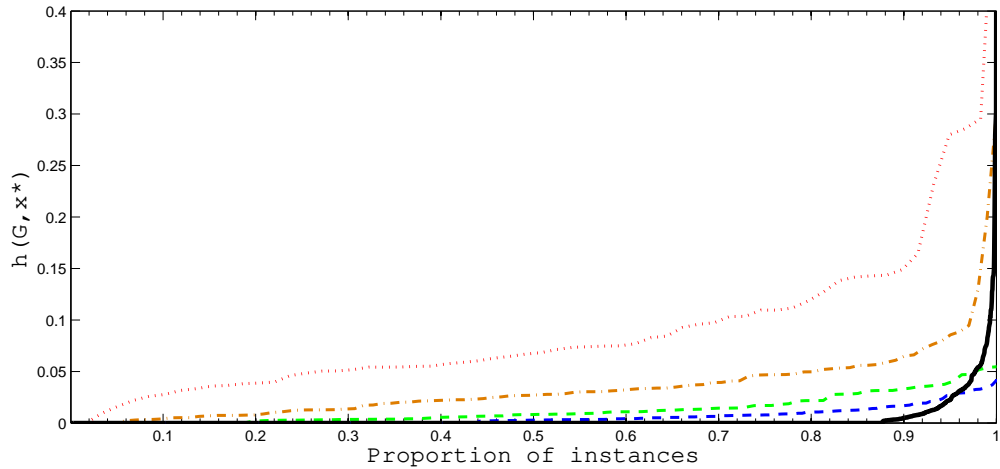}
\caption{\label{fig:D2_x0} The survival curve of the dataset $D_2$. Here the heuristic solution is stable ($h(G,x^*)=0$) for over 85\% of the instances.}
\end{center}
\end{figure*}

We sorted the instances according to $h(G,x^*)$, and plotted a survival graph showing the percentage of instances for which $h(G,x^*)<h$ (Figures~\ref{fig:D1_x0} and \ref{fig:D2_x0}). The following trends are apparent from the graphs:
\begin{enumerate}
	\item For most instances, $h(G,x^*)=0$. That is, $(P^*,x^*)$ is  PSPE. 
	\item For roughly $95\%$ of the instances in $D_1$ ($97\%$ in $D_2$), $h(G,x^*)\leq 0.05$. 
	\item As the gap $d=w(S_{i_+})-w(S_{i_-})$ increases, $h(G,x^*)$ also increases.
	\item In $D_2$, where the types are more diverse, $h(G,x^*)$ is lower, i.e. the heuristic solution is more stable. However if we condition on the gap $d$ (see dashed lines), then there is no significant difference. 
\end{enumerate}
It seems therefore that the diversity of types is responsible mainly for the reduction in  the average gap $d$, which in turn explains the improvement of the heuristic solution.

\subsection{Experimental profit distribution for three firms}
\label{sec:emp_revenue}
With more than two firms, a  PSPE is not guaranteed to be exist. Moreover, even when such a PSPE does exist, it will rarely be proportional (unless there happens to be an almost-balanced partition). We study the average case behavior by generating random instances and solve them as explained in the previous section. Thus we get the maximal and minimal share of the profit that firms may keep, and compare it with our baseline estimation of $r_0 = v(q)- z_0 \cdot q$, where $q=\floor{\sum_j w_j / k}$ and $z_0$ defined as above.

For every instance we computed the minimal and maximal revenue for firms in equilibrium, under the chosen partition (recall that in anonymous games, the profit of all firms is the same). Since the revenue of each firm $r$ is proportional to $SW(P^*) -  \max\{\sum_{j\in N}x_j\}$, in order to minimize $r$ we should maximize $x(N)$ and vice versa.

\begin{figure*}[p]
\begin{center}
\includegraphics[width=\textwidth]{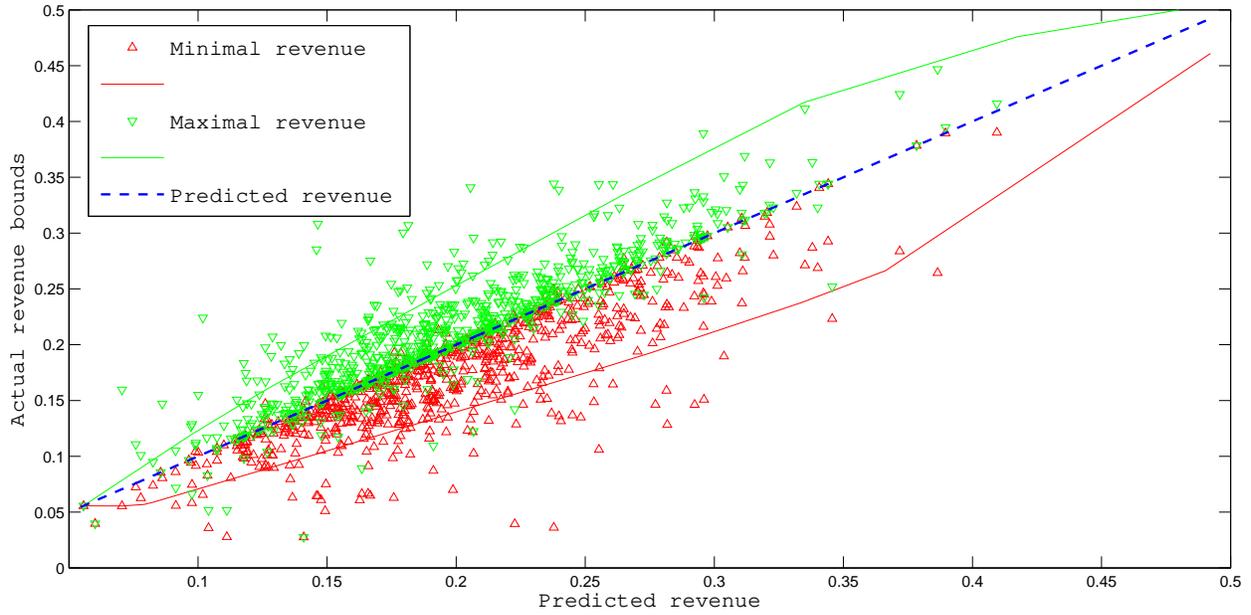}
\caption{\label{fig:D1_revenue} 
Actual revenue (minimal and maximal) w.r.t the predicted value $r_0$, for 800 random instances from $D_1$. The solid lines demarcate the area containing $90\%$ of the total 3000 samples for  which the prediction is most accurate. 
}
\end{center}
\end{figure*}

\begin{figure*}[p]
\includegraphics[width=\textwidth]{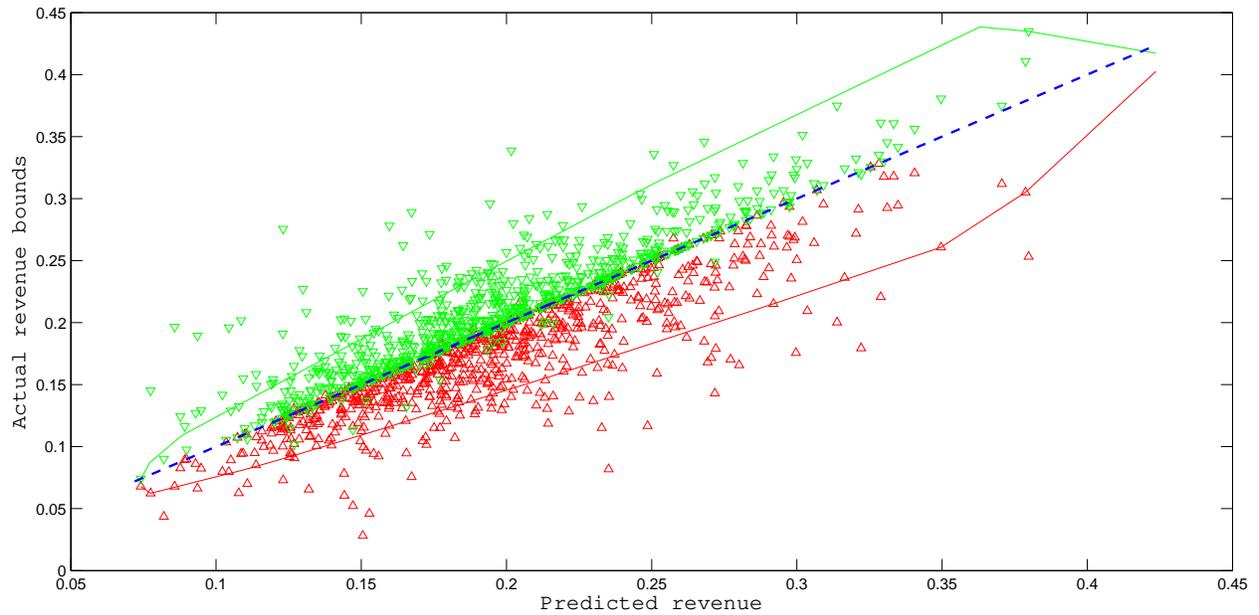}
\caption{\label{fig:D2_revenue}Revenue distribution in instances from $D_2$.}
\end{figure*}

\begin{figure*}[p]
\begin{center}
\includegraphics[width=\textwidth]{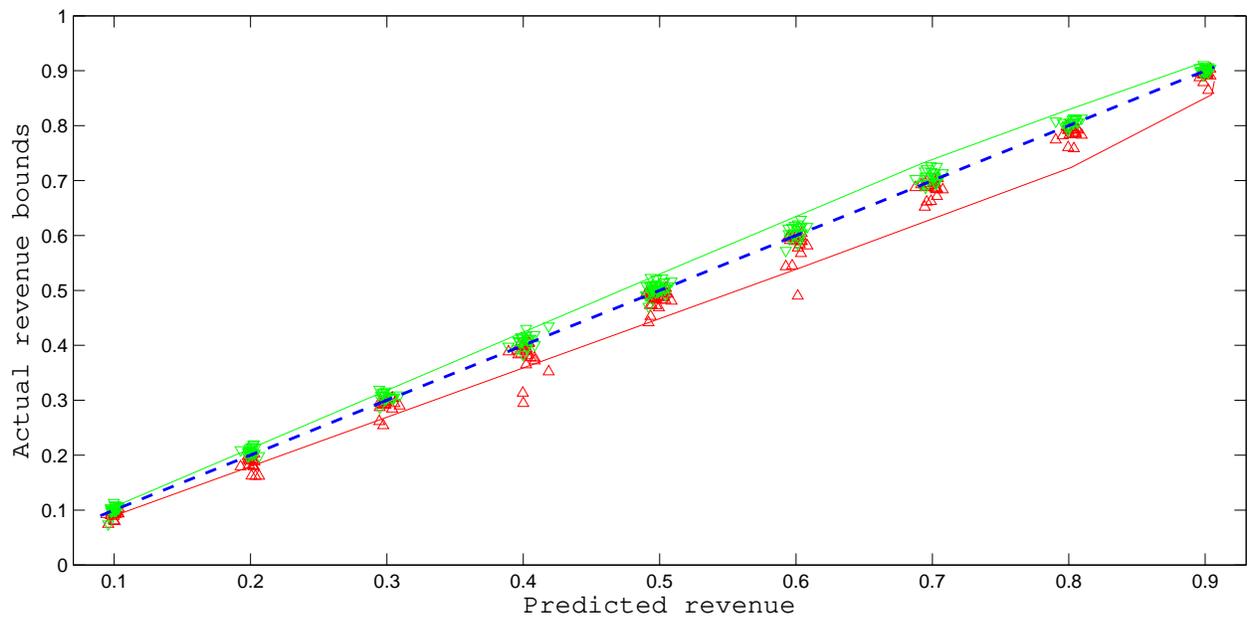}
\caption{\label{fig:D3_revenue}Revenue distribution in instances from $D_3$ (where $v(w)=w^\alpha$). We can see that the outcomes are clustered into 9 groups, each for one value of $\alpha$ (the leftmost corresponds to $\alpha=0.9$).}
\end{center}
\end{figure*}

Figures~\ref{fig:D1_revenue}, \ref{fig:D2_revenue}, and \ref{fig:D3_revenue} display how well the baseline $r_0$ predicts the actual revenue that firms can make in equilibrium. We can see that for the value function $v(w)=w^\alpha$, the actual revenue is quite close to $r_0=1-\alpha$, regardless of workers' weights (Fig.~\ref{fig:D3_revenue}). When the value function is random, we have a wider spread between the minimal and the maximal revenue of many instances. Still, the baseline $r_0$ is almost always in this range.

An interesting observation is that the \emph{existence} of a PSPE depends almost entirely on the weight vector $\vec w$, and not on the value function $v$ (as long as it is subadditive). In contrast, as can be seen in Figure~\ref{fig:D3_revenue}, the \emph{profit distribution} between firms and workers largely depends on the curvature of $v$, but almost not at all on the weights of workers.


\end{document}